\documentclass[traditabstract]{aa}

\usepackage{amsmath}
\usepackage{txfonts}
\usepackage{graphicx}
\usepackage{color}
\usepackage{xspace}
\usepackage{natbib}
\usepackage{lscape}
\usepackage{url}
\usepackage{refcount}
\graphicspath{{./figures/}}

\bibpunct{(}{)}{;}{a}{}{,} 

\begin{document}
\title{\textsl{Chandra} X-Ray Spectroscopy of the Focused Wind in the
Cygnus~X-1 System} 
\subtitle{II. The Nondip Spectrum in the Low/Hard State -- Modulations
  with Orbital Phase} 
\author{Ivica~Mi\v{s}kovi\v{c}ov\'a\inst{1} \and
        Natalie~Hell\inst{1,2} \and
        Manfred~Hanke\inst{1} \and        
        Michael~A.~Nowak\inst{3} \and
        Katja~Pottschmidt\inst{4,5} \and
        Norbert~S.~Schulz\inst{3} \and
        Victoria~Grinberg\inst{1,3} \and    
        Refiz~Duro\inst{1,6} \and 
        Oliwia~K.~Madej\inst{7,8} \and
        Anne M. Lohfink\inst{9} \and        
        J\'er\^{o}me Rodriguez\inst{10} \and
        Marion~Cadolle~Bel\inst{11} \and
        Arash~Bodaghee\inst{12} \and
        John~A.~Tomsick\inst{13} \and
        Julia~C.~Lee\inst{14} \and
        Gregory~V.~Brown\inst{2} \and        
        J\"orn~Wilms\inst{1}
}

\institute{Dr.~Karl Remeis-Sternwarte and Erlangen Centre for
  Astroparticle Physics, Universit\"at Erlangen-N\"urnberg,
  Sternwartstr.~7, 96049 Bamberg,
  Germany\thanks{joern.wilms@sternwarte.uni-erlangen.de} 
\and
  Lawrence Livermore National Laboratory, 7000 East Ave., Livermore,
  CA 94550, USA 
\and 
  MIT Kavli Institute for Astrophysics and Space
  Research, NE80, 77 Mass.\ Ave., Cambridge, MA 02139, USA 
\and
  CRESST, University of Maryland Baltimore County, 1000 Hilltop
  Circle, Baltimore, MD 21250, USA 
\and 
  NASA Goddard Space Flight Center, Astrophysics Science Division,
  Code 661, Greenbelt, MD 20771, USA 
\and
  AIT Austrian Institute of Technology GmbH, Donau-City-Str.~1, 1220
  Vienna, Austria
\and
  Department of Astrophysics/IMAPP, Radboud University
  Nijmegen, P.O.\ Box 9010, 6500 GL Nijmegen, The Netherlands 
\and 
  SRON Netherlands Institute for Space Research, Sorbonnelaan 2,
  3584 CA  Utrecht,The Netherlands 
\and 
  Institute of Astronomy, University of Cambridge, Madingley Road,
  Cambridge CB3 0HA, United Kingdom
\and 
  Laboratoire AIM, UMR 7158, CEA/DSM-CNRS-Universit\'e Paris Diderot,
  IRFU/SAp, 91191 Gif-sur-Yvette, France 
\and 
  Max Planck Computing and Data Facility, Gie\ss{}enbachstr.~2, 85748 Garching, Germany
\and
  Department of Chemistry, Physics, and Astronomy, Georgia College \&
  State University, Milledgeville, GA 31061
\and 
  Space Sciences Laboratory, 7 Gauss Way, University of California,
  Berkeley, CA 94720-7450, USA 
\and 
  Harvard John A. Paulson School of Engineering and Applied Sciences,
  and Harvard-Smithsonian Center for Astrophysics, 60 Garden Street
  MS-6, Cambridge, MA 02138, USA}

\authorrunning{Mi\v{s}kovi\v{c}ov\'a et al.}
\titlerunning{\textsl{Chandra} Spectroscopy of the Focused Wind of
  Cygnus X-1.\ II.}

\date{Received  / Accepted}

\abstract{The accretion onto the black hole in the system
  HDE~226868/Cygnus~X-1 is powered by the strong line driven stellar
  wind of the O-type donor star. We study the X-ray properties of the
  stellar wind in the hard state of Cyg~X-1 as determined with data
  from the \textsl{Chandra} High Energy Transmission Gratings. Large
  density and temperature inhomogeneities are present in the wind,
  with a fraction of the wind consisting of clumps of matter with
  higher density and lower temperature embedded in a photoionized gas.
  Absorption dips observed in the light curve are believed to be
  caused by these clumps. This work concentrates on the non-dip
  spectra as a function of orbital phase. The spectra show lines of
  H-like and He-like ions of S, Si, Na, Mg, Al and highly ionized Fe
  (\ion{Fe}{xvii}--\ion{Fe}{xxiv}). We measure velocity shifts, column
  densities, and thermal broadening of the line series. The excellent
  quality of these five observations allows us to investigate the
  orbital phase dependence of these parameters. We show that the
  absorber is located close to the black hole. Doppler shifted lines
  point at a complex wind structure in this region, while emission
  lines seen in some observations are from a denser medium than the
  absorber. The observed line profiles are phase dependent. Their
  shapes vary from pure, symmetric absorption at the superior
  conjunction to P~Cygni profiles at the inferior conjunction of the
  black hole. }


\maketitle

\section{Introduction}\label{sec:intro}
In the fifty years of persistent X-ray activity since its discovery in
1964 \citep{bowyer65}, the High-Mass X-ray Binary \object{Cygnus X-1}
(Cyg\,X-1) has become one of the best known X-ray sources, but there
are still many open questions even in regard to one of its most basic
properties, the nature of the accretion process. The system consists
of the supergiant O9.7~Iab type star HDE~226868 \citep{walborn73} and
a compact object in a 5.6\,d orbit
\citep{webster72,brocksopp99,gies03} with an inclination of
$i=27\fdg{}1\pm0\fdg{}8$ \citep{orosz11}. The system has a distance
$d=1.86^{+0.12}_{-0.11}$\,kpc \citep{Xiang11a,reid11}. Based on these
measurements, masses of $M_2=19.2\pm1.9M_\odot$ for the companion star
and $M_1=14.8\pm1.0\,M_\odot$ for the compact object have been deduced
\citep{orosz11}.

With a mass loss rate of
$\sim$$10^{-6}\,M_{\odot}\,\mathrm{year}^{-1}$ \citep{herrero95},
HDE~226868 shows a strong wind. Such winds are driven by radiation
pressure, due to copious absorption lines present in the ultraviolet
part of the spectrum on material in the stellar atmosphere
(line-driven or \citeauthor*{castor75}, \citeyear{castor75}, [CAK]
wind model), can reach very high velocities
\citep[$v_{\infty}\gg2000\,\mathrm{km}\,\mathrm{s}^{-1}$;][]{muijres12},
and are only produced by hot, early type O- or B-stars. Simulations
show that a steady solution of line-driven winds is not possible,
i.e., perturbations are present in the wind \citep{feldmeier97},
causing variations of density, velocity and temperature, which
compress the gas into small, cold, and overdense structures, often
referred to as ``clumps'' \citep[][and references
  therein]{oskinova12,sundqvist2013}. The existence of clumps is
supported by observations of transient X-ray absorption dips (lower
flux) in the soft X-ray light curves of such systems. \citet{Sako99}
estimate that more than 90\% of the total wind mass in
\mbox{Vela\,X-1} is concentrated in clumps, while the ionized gas
covers more than 95\% of the wind volume. Clumpiness with a filling
factor of 0.09--0.10 has been confirmed in Cyg\,X-1 by
\cite{rahoui11}.

Strong tidal interactions between the star and the black hole and
centrifugal forces make the wind distorted and asymmetric. Both the
wind density and the mass loss rate are enhanced close to the binary
axis, creating a so-called \emph{``focused wind''} \citep{friend82}.
Evidence for the presence of such a wind around HDE~226868, which
fills more than $\sim$90\% of its Roche volume \citep{giesbolton86b},
is the strong modulation of the \ion{He}{ii} $\lambda4686$ emission
line with orbital phase \citep{giesbolton86b} and the strong phase
dependence of other optical absorption lines, which are deepest around
orbital phase $\phi_\mathrm{orb}=0$ \citep{gies03}, i.e., during the
superior conjunction of the black hole. Further evidence comes from
the modelling of the IR continuum emission \citep{rahoui11}. Finally,
in the X-rays, the overall absorption column density and dipping vary
strongly with orbital phase and are largest at
$\phi_\mathrm{orb}\sim0.0$
\citep[e.g.,][]{li74,mason74,parsignault76,pravdo80,remillard84,kitamoto84,kitamoto89a,wen99,balucinska00,feng02,lachowicz06,poutanen08,hanke09},
consistent with the focused wind picture
\citep[e.g.,][]{li74,remillard84,balucinska00,poutanen08}. Due to the
presence of the focused wind, the accretion process in Cyg~X-1 is not
primarily wind accretion like in other HMXBs such as
\object{Vela~X-1}, \object{SMC X-1}, or \object{4U\,1700$-$37}
\citep{bondihoyle44,blondin95a,blondin94,blondin91}, but a small
accretion disk is present. Evidence for this disk comes from
detections of the disk's thermal spectrum
\citep[][]{balucinskachurch1995a,priedhorsky1979a}, an X-ray disk
reflection component, and a strong and broad Fe K$\alpha$ line
\citep[][and references therein]{tomsick14,duro11}.

Black hole binaries show two characteristic behaviors, called the
low/hard and the high/soft state. These states differ in the shape of
the X-ray spectrum, the timing properties, and the radio emission
\citep[and references therein]{fender99,belloni04,wilms06,belloni10}.
The spectrum of Cyg\,X-1 in the hard state is well described by a hard,
exponentially cut-off broken powerlaw with photon index
$\Gamma\sim1.7$ \citep{wilms06}. During the soft state, the power law
is steeper ($\Gamma\sim2.5$) and a luminous and less variable thermal
disk component appears \citep{wilms06}.

Cyg\,X-1 is often considered to be a hard state source, as it spent most
of the time in the hard state \citep{grinberg13a}. Transitions into
the soft state are observed every few years and ``failed state
transitions'', where the soft state is not entirely reached, are also
possible \citep{pottschmidt03}. Since 2010 the source's behavior has
been rather unusual and it has spent significantly more time in the
soft state. See \citet{grinberg13a} for a discussion of the long-term
changes in Cyg~X-1 from 1996 until the end of 2012.

Since X-rays from the black hole propagate through the stellar wind,
we can use these X-rays to probe its structure. For such a study it is
essential that the source is in the hard state or the
hard-intermediate state. The strong X-ray emission during the soft
state is sufficient to completely photoionize the stellar wind. As a
consequence, the wind is suppressed because the radiative driving
force of the UV photons from the donor star is reduced as the ionized
gas is transparent for UV radiation.

\begin{figure}
\resizebox{\hsize}{!}{\includegraphics{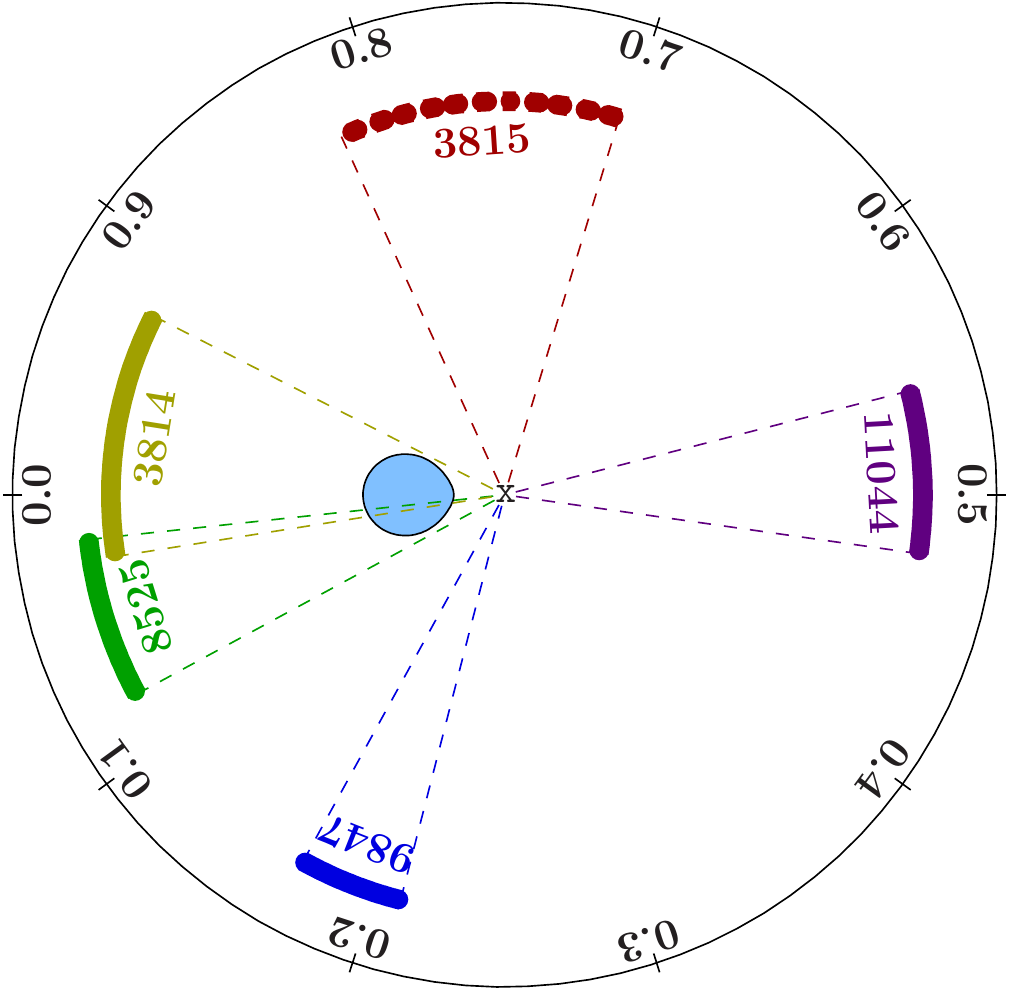}}
\caption{Orbital phase coverage of the \textit{Chandra} observations of Cyg\,X-1
  in the hard state analyzed in this work. Full arcs (dashed arcs)
  display TE mode (CC mode; for explanation see
  Sect.~\ref{sec:data_reduction}) observations. The labels correspond
  to the \textit{Chandra} ObsIDs. Phase $\phi_\mathrm{orb}=0$ corresponds to the
  superior conjunction of the black hole.  }
\label{chandra_coverage}
\end{figure}

In this paper, we extend earlier work on \textit{Chandra}
high-resolution grating spectra from Cygnus X-1 studying the ionized
material of the stellar wind of HDE~226868 during the low/hard state
\citep[][hereafter paper~I]{hanke09}. We perform a detailed study of
four observations at phases $\phi_\mathrm{orb}\sim0.05$, $\sim$0.2,
$\sim$0.5, and $\sim$0.75, and combine it with previous results of the
observation at $\sim$0.95 (paper~I). These data provide a unique set
of observations that allows us to probe all prominent parts of the
wind of Cyg\,X-1 (Fig.~\ref{chandra_coverage}). Our aim is to describe
the complex structure and the dynamics of the stellar wind.
Sect.~\ref{sec:obs} summarizes observations and data used in the
analysis. In Sect.~\ref{sec:continuum} and
Sect.~\ref{sec:modulations}, we present the data analysis and results
related to the continuum fitting and the H-like and He-like absorption
lines observed in the non-dip spectra. In Sect.~\ref{sec:variability},
we discuss their modulation with orbital phase.
Sect.~\ref{sec:line-profiles} discusses line profile variations and
results from plasma density diagnostics. We summarize our conclusions
in Sect.~\ref{sec:summary}. Overview tables and plots of all full
range spectra are given in appendix~\ref{sec:pars_and_spectra}.
Technical issues related to the analysis are discussed in
appendices~\ref{sec:order-sorting} and~\ref{sec:telemetry}.

\begin{figure*}\centering
\includegraphics[width=17cm]{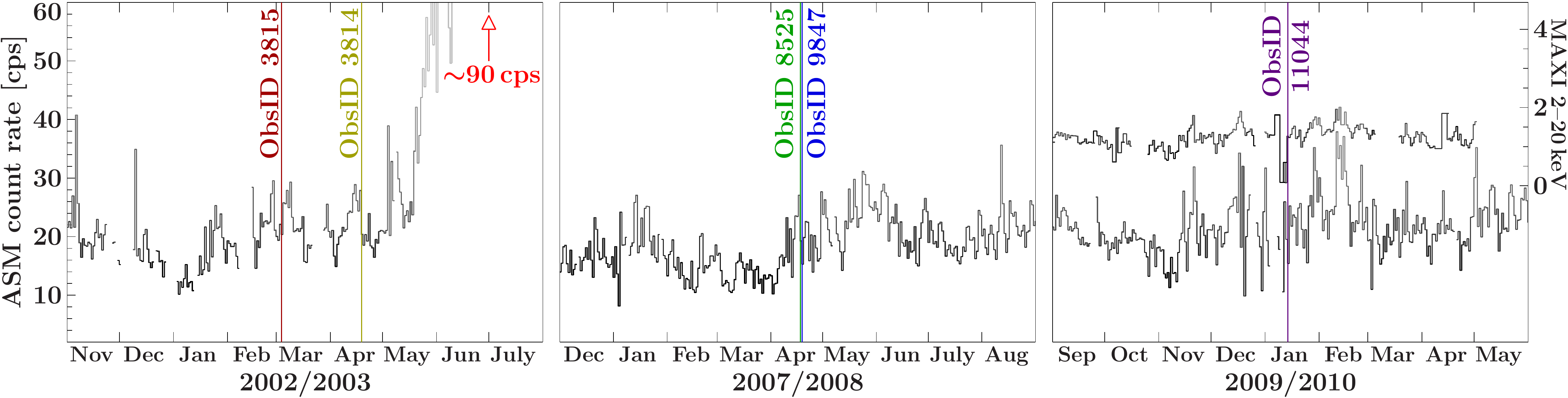}
\caption{1.5--12\,keV \textit{RXTE}-All-Sky Monitor count rate of Cyg\,X-1 with
  one day binning in the intervals of nine months during 2003, 2008
  and 2010 centered at the times of \textit{Chandra} observations: ObsIDs~3815
  and 3814 (\emph{left}), 8525 and 9847 (\emph{middle}), and 11044
  (\emph{right}), marked by vertical lines. The right panel also shows
  the MAXI light curve. }\label{asm}
\end{figure*}

\begin{figure*}\centering
\includegraphics[width=17cm]{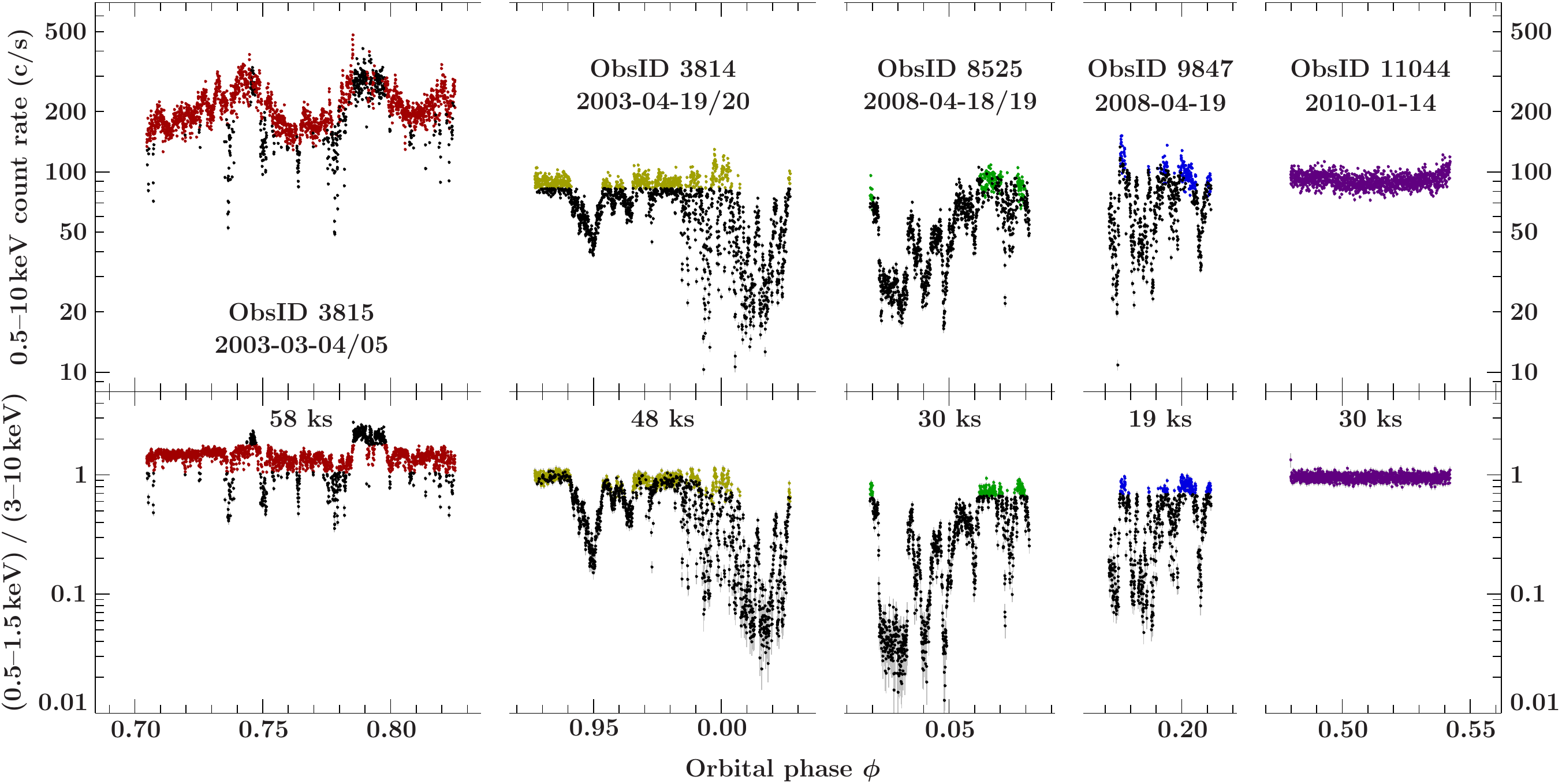} \caption{\emph{Upper
  panel}: Lightcurves of all five observations as a function of
  orbital phase. \emph{Lower panel}: variation of the hardness ratio.
  Except for the somewhat softer hard-intermediate state observation
  ObsId~3815, all observations were in the low/hard state. Dips are
  strongest at $\phi_\mathrm{orb}\sim0.0$, still present at
  $\phi_\mathrm{orb}\sim0.2$ and $\phi_\mathrm{orb}\sim0.75$, and they
  vanish at $\phi_\mathrm{orb}\sim0.5$. Colors indicate the parts of
  the observations used in the analysis (see
  Fig.~\ref{chandra_coverage}), while data in black were excluded from
  the analysis.}\label{lightcurves}
\end{figure*} 

\begin{table*}
\caption{Log of hard-state \textit{Chandra}-HETGS observations of Cyg\,X-1 used in
this paper.}\label{observations}

\centering

\begin{tabular}{ccccccccccc}
\hline
\hline
 & \multicolumn{2}{c}{Start date} & & & & \multicolumn{2}{c}{Count
 rates} & \multicolumn{3}{c}{Non-dip spectrum}\\
 ObsID & Date        & MJD & Mode & $T_\mathrm{exp}$ & $\phi_\mathrm{orb}$ & $r_\mathrm{ASM}$ & $r_\mathrm{\textit{Chandra}}$ & $h$ & $T_\mathrm{exp,non-dip}$
 & $L_\mathrm{0.5-10\,keV}$\\
       & yy-mm-dd  &     &      &  [ks]          &         & [cps]
       & [cps] & & [ks] & [$10^{37}$\,erg/s]\\
\hline                                                                                                      
3815 & 03-03-04 & 52702 & CC/g & 58.4 & 0.70--0.82 & 25.8 & 210.5 & 1.05--1.8 & 45 & 0.67\\
3814 & 03-04-19 & 52748 & TE/g & 48.3 & 0.93--0.03 & 17.5 & 88.6 & -- & 16.1 & 0.39\tablefootmark{a}\\
8525 & 08-04-18 & 54574 & TE/g & 30.1 & 0.02--0.08 & 14.2 & 87.5 & 0.667--0.94 & 4.4 & 0.43\\
9847 & 08-04-19 & 54575 & TE/g & 19.3 & 0.17--0.21 & 18.0 & 100.9 & 0.692--1.0 & 4.4 & 0.49\\
11044 & 10-01-14 & 55210 & TE/g & 30.1 & 0.48--0.54 & 18.4 & 91.1 & -- & 30.1 & 0.41\\ 
\hline
\end{tabular}

\tablefoot{Mode: CC/g -- Continuous Clocking, graded; TE/g -- Timed
  Exposure, graded; $T_\mathrm{exp}$: exposure time;
  $\phi_\mathrm{orb}$: orbital phase according to the ephemeris of
  \citet{gies03}; $r_\mathrm{ASM}$: \textit{RXTE}-ASM (1.5--12\,keV)
  count rate, averaged over the observation;
  $r_\mathrm{\textit{Chandra}}$: \textit{Chandra} spectral (non-dip)
  count rate; $h$ is the hardness ratio of the \textit{Chandra} first
  order count rates in the 0.5--1.5\,keV band to those in the
  3--10\,keV band; $T_\mathrm{exp,non-dip}$ --
  exposure time after exclusion of dips. Note that ObsID~8525 and~9847
  were scheduled as part of a multi-satellite campaign onto the
  system. For technical reasons this 50\,ksec observation was split
  into two parts. $L_\mathrm{0.5--10\,keV}$ -- source luminosity in
  the 0.5--10\,keV energy range, derived from the non-dip spectra
  assuming a source distance of 1.86\,kpc \citep{Xiang11a,reid11}.
  ${}^{a}$ \citet{hanke09}, corrected for the newer source distance.}
\end{table*}

\section{Observations and data reduction} \label{sec:obs}

\subsection{Selection of Observations}
The main purpose of this paper is to study the variation of the X-ray
spectrum with orbital phase during the hard state. Only six of the 17
available \textsl{Chandra}-HETG observations meet this condition:
ObsIDs 2415, 3814, 3815, 8525, 9847, and 11044 (see
Sect.~\ref{sec:state} for a discussion of the state classification).
All other \textsl{Chandra} observations of Cyg\,X-1 either caught the
source in the high/soft state or were too short to obtain spectra of a
sufficient signal-to-noise ratio. ObsID~2415, taken during the
intermediate state at $\phi_\mathrm{orb}\sim0.76$ was already analyzed
by \citet{miller05}. Table~\ref{observations} gives a log of the
remaining observations studied in this paper and
Fig.~\ref{chandra_coverage} depicts their orbital coverage. Results on
ObsID~3814 from paper~I will be combined with these new results in
Sect.~\ref{sec:modulations} and~\ref{sec:variability}.

To gauge the quality of the continuum modeling, we use simultaneous
pointed \textit{RXTE} observations (Sect.~\ref{sec:cont}), which were
reduced using our standard procedures as described, e.g., by
\citet{wilms06} or \citet{Grinberg_2015a}.

\subsection{Source State}\label{sec:state}
Figure~\ref{asm} shows the \textit{RXTE} All-Sky-Monitor
\citep[ASM;][]{levine96} lightcurve of Cyg\,X-1 around the times of the
\textsl{Chandra} observations. The average daily ASM data point
towards similar conditions during the first four observations.
According to the scheme of \citet{grinberg13a}, all four observations
are found in the hard state, with low countrate (14--18\,cps) in
ObsIDs~3814, 8525 and 9847 and a higher countrate of 26\,cps in
ObsID~3815.

ObsID~11044 was taken during the time when ASM was deteriorating
\citep{vrtilek12a,grinberg13a}. We therefore used MAXI data
\citep{matsuoka09a} for the assessment of this observation.  Six MAXI
measurements are simultaneous with ObsID~11044: they all show the
source in the hard state as defined by \citet{grinberg13a}.

\subsection{Data Reduction}\label{sec:data_reduction}

For the spectral analysis we used the first order spectra of
\textsl{Chandra}'s high and medium energy gratings \citep[HEG,
MEG;][]{canizares05}. Most HETG observations of Cyg\,X-1 in the analysis
were performed in timed exposure (TE) mode of the Advanced CCD imaging
spectrometer \citep[ACIS;][]{garmire03}. In this mode data are
nominally accumulated for a 3.2\,s frame time before being transferred
into a framestore and read out, increasing the probability of pile up.
In the cases of ObsIDs~3814, 8525, 9847, and 11044, the frame time was
reduced to 1.7\,s by using a 512 row CCD subarray readout.

In even brighter cases, as in ObsID~3815, where Cyg\,X-1 was in the
hard-intermediate state, a continuous clocking (CC) mode was applied
to avoid pile up and conserve line features. Here the CCD rows are
read out continously, which reduces the exposure per row to 2.85\,ms
\citep{garmire03} and no effects from pile up are expected. The
application of CC-mode comes at the expense of one spatial dimension
as the image is then reduced to $1024\times1$ pixel frames and the
$y$-image dimension is lost. As a consequence, all image photons
including non-source dispersed photons such as from the source
scattering halo get collapsed into the spectrum, order sorting becomes
more difficult, and calibration uncertainties are higher leading to
lesser determined continua. Some of these technical issues are
discussed in detail in appendix~\ref{sec:order-sorting}.

The data were processed with the standard Chandra Interactive Analysis
of Observations (CIAO) software, version~4.2. Further analysis was
done with the Interactive Spectral Interpretation System (ISIS),
versions 1.6.1 and 1.6.2 \citep{houck00}. Cross-sections were taken
from \citet{verner96}, abundances from \citet{wilms00}, and atomic
data from the Atomic Database, AtomDB v.~2.0.1 \citep{foster12}. Due
to the very low \textit{Chandra} background (compared to the source),
no background was subtracted from the final spectra. Data were grouped
to a minimum signal-to-noise ratio of $\mathrm{S}/\mathrm{N}=10$.
Unless noted otherwise, all uncertainties are at the 90\% level for
one parameter of interest \citep[$\Delta\chi^2=2.71$;][]{lampton:76a}.

For spectra from observations performed in TE mode the effect of
pile-up has to be considered. In the first order spectra it causes a
pure reduction of count rate. It is stronger in the MEG spectra than
in the HEG spectra due to the lower dispersion and higher effective
area of the MEG. The apparent flux reduction is most significant near
2\,keV (6--7\,\AA) where the spectrometer has the largest efficiency and
the highest count rates are obtained. As described in greater detail
in paper~I, pile up in the gratings can be modelled in ISIS
using the nonlinear convolution model \verb|simple_gpile2|. This model
describes the reduction of the predicted source count rate,
$S(\lambda)$, by pile up as
\begin{equation}
S'(\lambda)= S(\lambda) \cdot 
       \exp\left(-\beta\cdot S_\mathrm{tot}(\lambda)\right)
\end{equation}
where the total count rate $S_\mathrm{tot}(\lambda)$ is estimated from
the effective area and the assumed (model) photon flux, and where the
scale parameter $\beta$ is a fit parameter. $S_\mathrm{tot}(\lambda)$
includes the contributions of the $1^\mathrm{st}$, $2^\mathrm{nd}$,
and $3^\mathrm{rd}$ order spectra at the detector location
corresponding to $\lambda$.

\section{Continuum Modeling}\label{sec:continuum}

\subsection{Overview}\label{subsec:lcs}
A detailed look at the light curves and spectra of the five hard state
observations allows us to probe the structure of the wind and its
modulation with orbital phase. As shown in Fig.~\ref{lightcurves},
dipping is clearly present for most of the orbit, but the dip
frequency seems to be phase dependent. Light curves around
$\phi_\mathrm{orb}\sim0.0$ are modulated by strong and complex
absorption dips. Dipping occurs already at $\phi_\mathrm{orb}\sim0.7$
and has not ceased at $\phi_\mathrm{orb}\sim0.2$. The dip events
become shorter and shallower as the black hole moves away from
superior conjunction, as expected given that the line of sight through
the densest regions of the (focused) stellar wind is longest for
$\phi_\mathrm{orb}\sim0.0$. The data at $\phi_\mathrm{orb}\sim0.5$,
which probe only the outer regions of the stellar wind, do not show
any dipping. This distribution of dipping is consistent with
theoretical predictions and observations that see the high-density
focused wind close to the binary axis at $\phi_\mathrm{orb}=0$. A
consequence is a high probability to see dipping events at
$\phi_\mathrm{orb}=0$ and a much smaller probability for dipping at
$\phi_\mathrm{orb}=0.5$ \citep{balucinska00,poutanen08,boroson10a}.

The main goal of this paper is to study the effects of the stellar
wind on the X-ray spectrum. The time intervals where the data are
distorted by dips need to be removed from the analysis. Properties,
dynamics, and origin of the dips will be discussed by \citet[][2014,
  in prep.]{hell13a}. Following the discussion of \citet{hanke08},
different stages of dipping can be defined based on different count
rate levels in the light curve, or on its dependence in the
color-color diagram, or on the softness ratio. According to
\citet{kitamoto84} and confirmed by the lightcurves of our
observations, dips can last from several seconds to more than
10\,minutes, especially around superior conjunction. We therefore
extracted light curves with a 25.5\,s resolution, except for
ObsID~3814 where 12.25\,s were used (paper~I), in order to be able to
also identify short dipping intervals. For ObsIDs~3815, 8525, and 9847
the selection of non-dip intervals was based on the hardness ratio
defined as
$(0.5\mbox{--}1.5\,\mathrm{keV})/(3\mbox{--}10\,\mathrm{keV})$. See
Table~\ref{observations} for the exact selection criteria, which vary
between observations due to differences in the continuum shape, and
the resulting non-dip exposure times. As shown in
Fig.~\ref{lightcurves}, the selection criteria work well for all
observations, with only a small contribution due to residual dips with
low $N_\mathrm{H}$ remaining in the lightcurves. Choosing slightly
different selection criteria shows that these residual dips do not
affect our analysis. For ObsID~3814, the count-rate based selection of
paper~I is used. Since no dips are present in ObsID~11044, the full
$\sim$30\,ks of exposure can be used in the analysis.

Residual dipping present in the ``non-dip'' spectra can influence our
fitting results. In order to gauge the influence of dips on these
spectra we relaxed the hardness criterion and also extracted a
spectrum of ObsID~8525 which includes moderate dipping and refitted
the continum. The contaminated spectrum was chosen very
conservatively, it corresponds to a hardness $\ge 0.449$ in the lower
panel of Fig.~\ref{lightcurves}. In the combined spectrum,
$N_\mathrm{H}$ changed by 15\% from $5.5\times
10^{21}\,\mathrm{cm}^{-2}$ to $6.3\times 10^{21}\,\mathrm{cm}^{-2}$.
Because of our much more conservative selection criteria, the
systematic error in our non-dip spectra is significantly smaller than
that. We estimate it to be of the same order of magnitude as the
statistical errors of the fits.

\subsection{Continuum Model} \label{sec:cont}

The extracted non-dip spectra are characterized by an absorbed
continuum onto which a large number of absorption lines are
superimposed. As we are not focusing on a physical interpretation of
the continuum, we describe it with a simple empirical model that is
flexible enough to give an accurate representation of the proper
continuum spectrum, i.e., an absorbed power-law. Cross-checks with
simultaneous broad-band spectra from \textit{RXTE} performed during
our \textsl{Chandra} observations show that our continuum parameters
are in reasonable agreement between the two satellites. We look at two
extreme observations, ObsID~3815 and~11044. In ObsID~11044 the
\textit{Chandra} spectrum is neither distorted by calibration
features, nor influenced by strong absorption lines. Fitting the
\textit{Chandra} continuum model given in Table~\ref{8525} to the
simultaneous 3--6\,keV PCA spectrum, leaving the normalization value
free to vary to take into account the well-known flux-cross
calibration issues between the PCA and other satellites
\citep{nowak11}, gives a reasonable $\chi^{2}_\textrm{red}=1.65$
($\chi^{2}/\mathrm{dof}=9.95/6$) with consistent photon indices
for the two instruments. While formally not a very good fit, the
ratio between the data and the model in the PCA shows deviations of
$<$1\% and is therefore consistent with the calibration uncertainty of
the PCA in this energy band \citep{jahoda06,shaposhnikov12}. The large
$\chi^2$ is therefore due to PCA calibration systematics. Note that
the energy band chosen, 3--6\,keV, represents the maximum overlap
between the PCA and the HETGS, we deliberately do not extend the PCA
data to higher energies because we are only interested in determining
how well the continuum is described by the model in the HETGS band. 

As a second example we consider the continuum of \textit{Chandra}
ObsID~3815, which had to be modeled using a complicated and
nonphysical continuum. A direct comparison of this model with the
contemporaneous \textit{RXTE}-PCA data is complicated by the fact that
these data were unfortunately taken during one of the deep dips in the
lightcurve. We therefore extracted a \textit{Chandra} spectrum from
the aforementioned dip and made a comparison between strictly
simultaneous data only. We modeled it with the same continuum model
that was also used for the non-dip continuum, giving
$A_\mathrm{PL}$=$1.38\pm0.04$, $\Gamma=1.40\pm0.03$,
$N_\mathrm{H}=0.33\pm0.01\times 10^{22}\,\mathrm{cm}^{-2}$ and a
$\chi^{2}_\textrm{red}=1.07$ ($\chi^{2}/\mathrm{dof}=10550/9855$).
Applying the same fit to the PCA, leaving $N_\mathrm{H}$ fixed at the
\textit{Chandra} value, leaving only the normalization a parameter,
gives $A_\mathrm{PL}$=$1.61\pm0.01$ and the
$\chi^{2}_{\textrm{red}}=2.34$ ($\chi^{2}/\mathrm{dof}=14.07/6$). The
ratio between the data and the model shows again deviations of
$\leq$1\%, however, the residuals suggest that adjustment of the slope
of power law would improve the fit. A fit with $\Gamma$ left free,
$A_\mathrm{PL}$=$1.72\pm0.06$, $\Gamma=1.45\pm0.02$, is consistent
with the original fit to within the error bars. The data-to-model
ratio deviations lie below 0.5\%. This best-fit has an unphysically
good $\chi^{2}_\textrm{red}=0.21$ ($\chi^{2}/\mathrm{dof}=1.04/5$),
indicating that the systematic error in the PCA has been
overestimated. Fitting the PCA data without applying a systematic
error gives $\chi^{2}_\textrm{red}=0.87$
($\chi^{2}/\mathrm{dof}=4.34/5$). See \citet{nowak11},
\citet{wilms06}, and \citet{gierlinski99} for physical continuum
descriptions.

Neutral absorption is modelled with the \texttt{TBnew} model
\citep[][paper~I, and
\url{http://pulsar.sternwarte.uni-erlangen.de/~wilms/research/tbnew}]{juett06}.
Compared to the absorption model of \citet{wilms00}, \texttt{TBnew}
contains a better description of the absorption edges and allows a
simple fitting of columns of individual elements \citep[see,
e.g.,][]{hankeLMC10}. In gratings data such an approach is possible
when a strong absorption edge is present in the spectrum. For the
wavelength range studied here, the most important edge is the Ne
K-edge. Where indicated below, we therefore fitted the column of
neutral Ne independently of $N_\mathrm{H}$. Although K-edges of S, Si,
Mg, and Na are also present, they are not as clearly visible in the
spectra and the abundances of these elements were fixed to their
interstellar values \citep{wilms00}. Taking into account pile up, the
adopted continuum shape was
\begin{equation}\label{abs_pow_model}
N_\mathrm{ph}(E)=\mbox{\texttt{simple\_gpile2}}\otimes(\mbox{\texttt{TBnew}}\times\mbox{\texttt{powerlaw}}).
\end{equation}
In the following, we briefly discuss the continuum properties for the
four observations modeled here. We refer to paper~I for
continuum description of the non-dip spectrum of ObsID~3814.

\begin{table}
\caption{Continuum parameters of ObsID 8525, 9847 and 11044}\label{8525}
\centering
\begin{tabular}{lllll}
 \hline
 \hline
Parameter & ObsID 8525 & ObsID 9847 & ObsID 11044 \\
             & $\phi_\mathrm{orb}\sim0.05$ & $\phi_\mathrm{orb}\sim0.2$ & $\phi_\mathrm{orb}\sim0.5$ \\
 \hline
 \multicolumn{4}{l}{\texttt{power-law}}\\
 $A_\mathrm{PL}$ [$\mathrm{s}^{-1}\,\mathrm{cm}^{-2}\,\mathrm{keV}^{-1}$] & $1.23^{+0.03}_{-0.02}$ & $1.45\pm0.02$ & $1.38\pm0.01$\\ 
 $\Gamma$ & $1.43^{+0.02}_{-0.01}$ & $1.45\pm0.01$ & $1.59\pm0.01$\\ 
 \multicolumn{4}{l}{\texttt{TBnew}}\\
 $N_\mathrm{H}$ [$10^{22}\,\mathrm{cm}^{-2}$] & $0.55^{+0.03}_{-0.02}$ & $0.47\pm0.02$ & $0.56\pm0.01$\\ 
 $N_\mathrm{Ne}$ [$10^{18}\,\mathrm{cm}^{-2}$] & $0.77\pm0.18$ &$1.13^{+0.14}_{-0.15}$ & $0.58\pm0.05$\\ 
 $N_\mathrm{Fe}$ [$10^{17}\,\mathrm{cm}^{-2}$] & -- & -- & $1.5\pm0.2$ \\ 
 \multicolumn{4}{l}{\texttt{simple\_gpile2}}\\
$\beta_{\mathrm{HEG}-1}$ & $4.40^{+0.26}_{-0.41}$ & $3.8^{+0.1}_{-0.2}$ & $3.8\pm0.1$\\
$\beta_\mathrm{HEG+1}$ &$4.74^{+0.26}_{-0.40}$  & $4.01\pm0.16$ & $4.08\pm0.08$\\
$\beta_{\mathrm{MEG}-1}$ & $6.01^{+0.15}_{-0.20}$ & $5.69\pm0.09$ & $5.46\pm0.04$\\
$\beta_\mathrm{MEG+1}$ & $6.41^{+0.13}_{-0.17}$ & $6.41\pm0.07$ & $6.15\pm0.03$\\
\hline  
$\chi^{2}$ & 3174.49 & 3661.78 & 12142.47 \\
$\mathrm{dof}$ & 2985 & 3382 & 11040 \\ 
$\chi^{2}_\mathrm{red}$ & 1.06 & 1.08 & 1.10\\ 
\hline
\end{tabular}
\tablefoot{$A_\mathrm{PL}$ -- flux density of the power-law at 1\,keV;
  $\Gamma$ -- photon index of the power-law; $N_\mathrm{H}$ --
  hydrogen column density, $N_\mathrm{Ne}$ and $N_\mathrm{Fe}$ --
  neutral column densities of Ne and Fe; \texttt{simple\_gpile2}
  $\beta$ -- pile-up scale parameters in units of $10^{-2}$\,s\,\AA.}
\end{table} 
 
\subsection{The Continuum of ObsIDs 8525 and 9847 ($\phi_\mathrm{orb}\sim0.05$
  and $\phi_\mathrm{orb}\sim0.2$)}\label{sec:8525-cont} After filtering for dips
only $\sim$4.4\,ks of non-dip data remain for each of the two
observations. In order to stay above a signal-to-noise ratio of 10,
the continua of these two observations were modeled in the wavelength
range 2\,\AA--15\,\AA\ for MEG and 2\,\AA--12\,\AA\ for HEG. Best-fit
parameters are listed in Table~\ref{8525}.

Both observations show many short, but strong, dips in the light
curve. It is probable that even after the exclusion of dips, the
non-dip spectrum is contaminated by faint dips, which may have an
influence on parameters obtained in the analysis.

\subsection{The Continuum of ObsID 11044
  ($\phi_\mathrm{orb}\sim0.5$)}\label{sec:11044-cont}

The high signal to noise ratio of this $\sim$30\,ks observation allows
us to model the continuum in the range of 1.7\,\AA--20\,\AA. The
continuum is well described by Eq.~\eqref{abs_pow_model}. The Ne- and
also Fe-column densities were allowed to vary and are mainly
constrained by the Ne K and Fe $\mathrm{L}_2$ / $\mathrm{L}_3$ edges
at 14.3\,\AA{} and 17.2\,\AA/17.5\,\AA, respectively, which are
very prominent in the spectrum. Best fit parameters are again shown in
Table~\ref{8525}. As also indicated by similar behavior in
\textsl{RXTE}-ASM (Fig.~\ref{asm}), ObsID~11044 was performed in a
similar state as ObsID~8525 ($\phi_\mathrm{orb}\sim0.05$) and~9847
($\phi_\mathrm{orb}\sim0.2$), and therefore it is not surprising that the
spectral parameters are very similar.

\subsection{The Continuum of ObsID 3815
  ($\phi_\mathrm{orb}\sim0.75$)}\label{sec:3815-cont}

The analysis of the $\sim$45\,ks non-dip spectrum of this CC-mode
observation is complicated by the lack of imaging information and by
calibration
issues\footnote{\url{http://cxc.harvard.edu/cal/Acis/Cal_prods/ccmode/ccmode_final_doc02.pdf}}.
Below 2\,\AA\ both spectra show an excess of up to
$50\,\mathrm{ph}\,\mathrm{cm}^{-2}\,\mathrm{s}^{-1}$\,\AA${}^{-1}$ for
HEG and up to
$150\,\mathrm{ph}\,\mathrm{cm}^{-2}\,\mathrm{s}^{-1}$\,\AA${}^{-1}$ in
the MEG. This excess is probably caused by contamination of the
spectra by the dust scattering halo surrounding the source, which is
clearly visible in the detector images of other observations
\citep[see also][]{Xiang11a}, or contamination from the wings of 0th
order image. As there is no imaging information available in CC-mode,
it is not possible to correct for this contamination. Since the low
signal-to-noise ratio of the data above 15\,\AA\ does not allow
detailed spectral modeling, only the 2--15\,\AA\ HEG and
2.5--15\,\AA\ MEG data are taken into account in the further analysis.

Unlike for the TE-mode data, the continuum here cannot be described by
the simple power law of Eq.~\eqref{abs_pow_model}, as non-physical
curvature caused by calibration issues is present in the spectrum. We
model this curvature by adding two non-physical Gaussian components,
centered at $\sim$1.12\,keV and $\sim$2.25\,keV. Calibration issues
causing slope differences between both instruments also necessitated
separate modeling of the continua of the HEG and the MEG. The final
parameters of the continuum fit are summarized in
Table~\ref{cont_3815}.

Note that despite the fact that the broad band CC mode calibration is
suboptimal, its relative calibration over small wavelength intervals
is still good. This means that parameters of absorption lines can
nevertheless be measured. For example, equivalent widths of narrow
absorption lines do not depend on the overall shape of the continuum
and are therefore not affected by local fitting or the shape of the
non-physical continuum model. Since the Ne column is, as in the other
observations, obtained mainly from modeling the Ne K-edge, i.e., a
local quantity, the Ne column density could be measured independently
from the total column.

\begin{table}
 \caption{Continuum parameters of ObsID 3815}\label{cont_3815}
\centering
 \begin{tabular}{ll@{\hspace*{1ex}}l@{\hspace*{1ex}}l@{\hspace*{1ex}}l}
  \hline
  \hline
   Parameter & HEG$-1$ & HEG+1 & MEG$-1$ & MEG+1 \\
  \hline
\multicolumn{5}{l}{\texttt{power-law}}\\
   $A_\mathrm{PL}$ & $2.70\pm0.01$ & $2.68\pm0.01$ & $2.24\pm0.01$ & $2.53\pm0.01$ \\ 
  $\Gamma$ & $1.81\pm0.01$ & $1.73\pm0.01$ & $1.64\pm0.01$ & $1.69\pm0.01$ \\
\multicolumn{5}{l}{\texttt{tbnew}}\\
   $N_\mathrm{H}$ [$\mathrm{cm}^{-2}$] & \multicolumn{4}{l}{$\left(0.352^{+0.005}_{-0.003}\right)\times10^{22}$} \\
   $N_\mathrm{Ne}$ [$\mathrm{cm}^{-2}$] & \multicolumn{4}{l}{$\left(0.76^{+0.01}_{-0.04}\right)\times10^{18}$} \\
\multicolumn{5}{l}{\texttt{gauss} $E\sim1.12$\,keV ($\lambda\sim11$\,\AA)} \\
  $A$ & 0.23 & 0.17 & 0.15 & 0.04 \\
    $\sigma$ & 0.178 & 0.163 & 0.159 & 0.099 \\
\multicolumn{5}{l}{\texttt{gauss} $E\sim2.25$\,keV ($\lambda\sim5.5$\,\AA)} \\
  $A$  & 0.03 & 0.01 & 0.07 & 0.04 \\
    $\sigma$ & 0.232 & 0.117 & 0.283 & 0.196 \\
 \hline
   $\chi^{2}/\mathrm{dof}$ & \multicolumn{4}{l}{29805/14406}\\
   $\chi^{2}_\mathrm{red}$ & \multicolumn{4}{l}{2.07}\\
\hline
 \end{tabular}
\tablefoot{$A_\mathrm{PL}$ -- flux density of the power-law at 1\,keV
  in $\mathrm{cm}^{-2}\mathrm{s}^{-1}\,\mathrm{keV}^{-1}$; 
$\Gamma$ -- photon index of the power-law;
$A$ -- flux of the broad line feature in
$\mathrm{ph}\,\mathrm{cm}^{-2}\,\mathrm{s}^{-1}$; 
$\sigma$ -- line width in keV. }
\end{table}

\section{Modeling of Line Features}\label{sec:modulations}
\subsection{Introduction}
Visual inspection of the non-dip spectra reveals that all spectra show
discrete line absorption due to highly ionized material, mostly from
H-like and He-like ions of S, Si, Al, Mg, Na, and Ne. We also observe
intercombination (i) and forbidden (f) emission lines of He-like ions.
Various L-shell transitions of Fe are also present. Table~4 of paper~I
gives a complete list of transitions from H- and He-like ions present
in the spectrum of ObsID~3814 ($\phi_\mathrm{orb}\sim0.95$). Most of these
absorption lines are also present in the spectra of ObsID~8525, 9847,
and 3815 ($\phi_\mathrm{orb}\sim0.0$--0.2 and $\phi_\mathrm{orb}\sim0.75$). Since the
wavelength range used here is smaller than in paper~I, lines
from \ion{O}{viii}, \ion{O}{vii}, \ion{Fe}{xxvi}, and \ion{Ni}{xxviii}
are outside of the investigated spectral range.

To describe the lines from the H- and He-like ions we fit all
transitions of the series from a given ion simultaneously using a
curve of growth approach (see paper~I for a detailed description of this
model). This approach allows us to describe weaker and
often blended lines, while their description with separate Gaussian
profiles would often be impossible. For each line series, the fit
parameters are the column density of the ion responsible for the line,
$N_i$, the Doppler shift, $v_i$, and the thermal broadening parameter,
$\xi_i$. All line shapes are modeled using Voigt profiles.

In addition to the lines from line series, other absorption and
emission lines are also present in the spectra. Where such lines were
identified, they were added by hand. Unless noted otherwise, line
shapes were described using Voigt profiles, with fit parameters being
the thermal broadening parameter, $\xi$, the natural line width,
$\Gamma$, i.e., the full width at half maximum of the Lorentzian
component of the Voigt profile, and the line flux, $A$. Negative
fluxes denote absorption lines.

The following sections discuss the details and peculiarities of each
observation.

\subsection{Line spectroscopy of ObsIDs 8525 and 9847
  ($\phi_\mathrm{orb}\sim0.05$, $\phi_\mathrm{orb}\sim0.2$)} \label{sec:8525_ion_gas}

The line identifications for these observations are shown in
Figs.~\ref{8525_spectrum} and~\ref{9847-spectrum}. While we are able
to fit all line series in ObsID~9847, \ion{Ar}{xvii} and \ion{Ca}{xix}
could not be constrained in ObsID~8525. In addition to the absorption
lines from H- and He-like ions and Fe, the spectra show evidence for
lower ionization absorption lines of \ion{Si}{xii} (Li-like),
\ion{Si}{xi} (Be-like), \ion{Si}{x} (B-like), and \ion{Si}{ix}
(C-like) in the 6.6\,\AA--7.0\,\AA\ band. As we will discuss in
greater detail in our analysis of the dip spectrum (Hirsch et al., 2016,
in prep.), these features become very strong during the deepest phases
of absorption dips. Their appearance in the non-dip spectrum therefore
reveals the presence of cooler dense material along the line of sight.
This is not surprising given that ObsIDs~8525 and~9847 represent the
densest part of the wind, $\phi_\mathrm{orb}\sim0.0$--0.2, where most
dipping is observed. The spectra taken farther away from the superior
conjunction, which are far less dominated by dips, are virtually free
of lower ionization Si features, with the exception of possible
detections of a \ion{Si}{xii} line in ObsID~11044
($\phi_\mathrm{orb}\sim0.5$) and a \ion{Si}{xi} line in ObsID~3815
($\phi_\mathrm{orb}\sim0.75$). We note, however, that in all cases the
optical depth of the lines is much smaller than that seen during dips
and consider our results to be representative of the non-dip spectrum.
Best-fit line parameters of the Si features are listed in
Table~\ref{low_si}. Note that \ion{Si}{xii} $\lambda$6.72\,\AA\ blends
with \ion{Mg}{xii}\,Ly$\gamma$ and the \ion{Si}{xiii}\,f emission
line, while \ion{Si}{xi} $\lambda$6.785\,\AA\ blends with a line from
\ion{Fe}{xxiv}. The parameters of these lines are therefore difficult
to constrain\footnote{Because of blends, the Si absorption lines
  visible in the non-dip spectrum of ObsID 3814 were not identified as
  such in Fig.~10 of paper~I.}.

\ion{Si}{xiii}\,forbidden line at 6.74\,\AA\ is a single emission
feature in the spectrum and its modeling is ambiguous due to possible
blends described above. Fixing $\Gamma$ and fitting the feature gives
line fluxes of
$0.29^{+0.15}_{-0.13}\times10^{-3}\,\mathrm{ph}\,\mathrm{s}^{-1}\mathrm{cm}^{-2}$
in ObsID~8525 and
$0.56\pm0.16\times10^{-3}\,\mathrm{ph}\,\mathrm{s}^{-1}\mathrm{cm}^{-2}$
in ObsID~9847, while the thermal broadening parameter, $\xi$, is
almost unconstrained.

\begin{table*}
 \caption{Low ionization Si lines in ObsIDs 8525 ($\phi_\mathrm{orb}\sim0.05$)
   and 9847 ($\phi_\mathrm{orb}\sim 0.2$)}\label{low_si}
\centering
\begin{tabular}{ccc cccc}
\hline
\hline
Line & $\lambda_\mathrm{lab}$ & ObsID & $\lambda_\mathrm{obs}$ & $A$                                                       & $\Delta v$                       & $\xi$\\
     & [\AA]                  &       & [\AA]                  & [$10^{-4}\,\mathrm{ph}\,\mathrm{s}^{-1}\mathrm{cm}^{-2}$] & [$\mathrm{km}\,\mathrm{s}^{-1}$] & [$\mathrm{km}\,\mathrm{s}^{-1}$]\\
\hline
\ion{Si}{xii} (Li) & $6.7200\pm0.0003$ & 8525 & $6.716\pm0.004$ & $-3.4^{+1.0}_{-1.2}$ & $-162\pm162$ & $130^{+320}_{-100}$\\
                   &                   & 9847 & $6.715\pm0.003$ & $-5.9\pm1.3$ & $-244\pm114$ & $380^{+200}_{-220}$\\
\ion{Si}{xi} (Be)  & $6.7848\pm0.0003$ & 8525 & line not present & \multicolumn{3}{c}{}\\
                   &                   & 9847 & $6.7750\pm0.0001$ & $-3.2^{+0.8}_{-0.9}$ & $-434\pm7$ & $15^{+334}_{-4}$\\
\ion{Si}{x} (B)    & $6.8565\pm0.0002$ & 8525 & $6.858\pm0.008$ & $-2.7^{+1.2}_{-1.3}$ & $43\pm332$ & $600^{+600}_{-500}$\\
                   &                   & 9847 & $6.859\pm0.001$ & $-1.2^{+1.3}_{-1.4}$ & $126\pm33$ & $\le2595$ \\
\ion{Si}{ix} (C)   & $6.9285\pm0.0003$ & 8525 & $6.9228\pm0.0001$ & $-11^{+11}_{-15}$ & $\-248\pm2$ & $\le749$\\
                   &                   & 9847 & line not present & \multicolumn{3}{c}{}\\
\hline
\end{tabular}
\tablefoot{$\lambda_\mathrm{lab}$: laboratory wavelength
  \citep{hell13a}, $\lambda_\mathrm{obs}$: observed wavelength, $A$:
  line flux (negative: absorption), $\Delta v$: velocity
  shift, $\xi$: thermal broadening parameter.} 
\end{table*}

\subsection{Line spectroscopy of ObsID 3815
  ($\phi_\mathrm{orb}\sim0.75$)} \label{sec:3815_ion_gas} 

The full range spectrum is given in Fig.~\ref{abs-lines}. In this
observation, the line series of \ion{Ar}{xvii}, \ion{Ca}{xix},
\ion{Ca}{xx} and \ion{Fe}{xvii} could not be fitted. The approach of
line series fitting does not allow us to describe every single line
perfectly, but this is mostly only the case for weak emission lines or
Fe blends. Notable is the discrepancy in the case of \ion{Ne}{x}. If
\ion{Ne}{x}\,Ly$\alpha$ (12.13\,\AA) is modeled properly, then
\ion{Ne}{x}\,Ly$\beta$ (10.24\,\AA), \ion{Ne}{x}\,Ly$\gamma$
(9.71\,\AA), and \ion{Ne}{x}\,Ly$\delta$ (9.48\,\AA) are predicted
to be weaker than observed. To find the reason for this discrepancy,
we excluded a very narrow region of the spectrum where the
\ion{Ne}{x}\,Ly$\alpha$ is located and fitted the spectrum without
this line. All other lines of the \ion{Ne}{x} series were described
very well. There is, however, no reason to assume that the data in the
region of \ion{Ne}{x}\,Ly$\alpha$ are of low quality, especially not
for non-dip spectra with their good statistics, and this discrepancy
is probably due to contamination by nearby Fe lines.
\ion{Ne}{x}\,Ly$\alpha$ also appears to be asymmetric, possibly
being an indication of P~Cygni profiles (see
Sect.~\ref{sec:11044_ion_gas}). \citet{miller05}\footnote{ObsID~2415:
  $\sim$32\,ks at $\phi_\mathrm{orb}\sim0.76$, CC mode, $\sim$\,twice as high
  flux as common \label{f:obs2415}} reported inconsistencies in
equivalent widths between individual lines in the \ion{Ne}{x} series.
Apart from (partial) saturation of the resonance line or blends of
lines, the authors argued that another possible explanation would be
that lines come from an inhomogeneous region, or from different
regions simultaneously. \citet{marshall01}\footnote{ObsID~1511:
  12.7\,ks at $\phi_\mathrm{orb}\sim0.84$, CC mode} described an opposite
problem, that while the Ly$\alpha$ lines were strong, no Ly$\beta$
lines were detected, concluding that the lines were not saturated and
that the absorbing gas covered more than $\sim$50\% of the source.

Emission lines are present in the spectrum but are generally much
weaker than the absorption lines. Since full width at half maximum,
$\Gamma$, and the thermal broadening, $\xi$, of the line are
correlated for Voigt profiles, it was difficult to constrain their
parameters (Table~\ref{emi-lines}). We therefore set $\Gamma$ to a
fixed value that is located at the intersection of the confidence
contours for all individual emission lines from the free fits, which
appeared to be split into two groups: for stronger lines, $\Gamma$ was
fixed to 15\,eV and the thermal broadening reaches values around
$\sim500\,\mathrm{km}\,\mathrm{s}^{-1}$. The full width at half
maximum of the weaker lines was fixed to $1.5\,\mathrm{eV}$ and $\xi$
is mostly consistent with zero.

\begin{table}
  \caption{Intercombination (i) and forbidden (f) emission lines from
    He-like ions at $\phi_\mathrm{orb}\sim0.75$
    (ObsID~3815)}\label{emi-lines}
\centering
\begin{tabular}{lcccc} 
\hline 
\hline
 & $\lambda$ & $A$ & $\xi$  \\ 
 & [\AA] & [$10^{-3}\,\mathrm{ph}\,\mathrm{s}^{-1}\mathrm{cm}^{-2}$] & [$\mathrm{km}\,\mathrm{s}^{-1}$]\\
\hline
\ion{Ne}{ix}\,f & 13.726 & $0.12\pm0.09$ & $\le200$ \\
\ion{Ne}{ix}\,i & 13.561 & $0.59^{+0.09}_{-0.10}$ & $140^{+70}_{-100}$ \\ 
\textbf{\ion{Na}{x}\,f} & 11.199 & $0.61^{+0.18}_{-0.31}$ & $590^{+370}_{-90}$ \\
\textbf{\ion{Na}{x}\,i} & 11.089 & $0.84^{+0.62}_{-0.19}$ & $1200^{+2600}_{-500}$ \\
\ion{Mg}{xi}\,f & 9.316 & $0.17^{+0.07}_{-0.06}$ & $\le200$ \\
\ion{Mg}{xi}\,i & 9.235 & $0.47\pm0.07$ & $\le180$ \\
\ion{Al}{xii}\,f & \multicolumn{3}{c}{--} \\
\textbf{\ion{Al}{xii}\,i} & 7.81 & $0.4\pm0.08$ & $550^{+200}_{-150}$ \\
\ion{Si}{xiii}\,f & 6.742 & $0.79^{+0.05}_{-0.07}$ & $<149$ \\
\ion{Si}{xiii}\,i & 6.685 & $0.27^{+0.08}_{-0.07}$ & $360^{+140}_{-310}$ \\
\ion{S}{xv}\,f & \multicolumn{3}{c}{--}\\
\ion{S}{xv}\,i & 5.054 & $0.05^{+0.09}_{-0.05}$ & $\le200$ \\
\hline 
\end{tabular}
\tablefoot{Full width at half maximum, $\Gamma$, was fixed to 15\,eV for the lines indicated in
  \textbf{bold}. For all the other lines, it was fixed to
  1.5\,eV. Emission lines of \ion{Ar}{xvii} and \ion{Ca}{xix}
  were not present in the spectrum.}
\end{table}

\subsection{Line spectroscopy of ObsID 11044
  ($\phi_\mathrm{orb}\sim0.5$)} \label{sec:11044_ion_gas}
Whereas resonance transitions were always detected in absorption at
early orbital phases, the spectrum at $\phi_\mathrm{orb}\sim0.5$, taken during
the inferior conjunction of the black hole, shows P~Cygni profiles
with an emission component at the rest wavelength and a weak
blue-shifted absorption component. Due to the complex profiles, in
this observation line series fitting is not feasible. Describing the
P~Cygni profiles as the sum of a positive and a negative Voigt profile
allows us to model the shapes of the Ly$\alpha$ transitions of
\ion{Si}{xiv}, \ion{Mg}{xii}, and \ion{Ne}{x}, \ion{Ne}{x}$\beta$
and $\delta$ quite well (Table~\ref{11044_p-cygni_my}). In addition,
the data show the \ion{Mg}{xi} triplet, four pure emission lines, and
only two pure absorption lines. See Fig.~\ref{11044-spectrum} for the
full spectrum.

Although the interpretation of P~Cygni profiles is challenging, we
attempt to determine column densities for the absorption part of the
profiles using \citep[][see also paper~I]{mihalas78}
\begin{equation}\label{eq:COGlin}
 N_i=\frac{mc^2 W_\lambda}{\pi e^2 \cdot f_{ij}\lambda_0^2} =
           \frac{1.13\times10^{17}\mathrm{cm}^{-2}}{f_{ij}} 
           \cdot \left(\frac{\lambda_0}{\mbox{\AA}}\right)^{-2} 
           \cdot \left(\frac{W_\lambda}{\mbox{m\AA}}\right),
\end{equation}
assuming that the lines are not saturated ($\tau(\nu)\ll1$). Here,
$W_\lambda$ is the equivalent width of the line, and $f_\mathrm{ij}$
is the oscillator strength of the transition, in this case a sum of
all transitions contributing to the absorption line. The values of
$N_\mathrm{i}$ are summarized in Table~\ref{11044_p-cygni_my}. Column
densities for \ion{Ne}{x}$\beta$ and $\delta$ are outliers. Values of
$\sim$75 and $\sim85\times10^{16}\,\mathrm{cm}^{-2}$ are far too
high given the column density of \ion{Ne}{x}$\alpha$. This deviation
is most likely due to a partial filling of the absorption component by
the strong emission wing, which cannot be determined reliably due to
the low signal-to-noise ratio of the line (Fig.~\ref{11044-spectrum}).
Columns were also determined for the absorption line at
$\lambda$\,6.705\,\AA\ assuming it is from Li-like \ion{Si}{xii},
using the oscillator strength of \cite{behar02}.

The fact that the columns of the absorption lines are much
smaller in this observation compared to earlier orbital phases
demonstrates that the column density of the ionized absorber is
strongly modulated with orbital phase, although the blending of the
absorption and the emission component strongly influences the
interpretation of the columns found for this observation. A more
detailed study would require fits of P~Cygni profile models using line
profile shapes in which the exciting source is not situated at the
center of the complex, asymmetric flow. To our knowledge, such
calculations are not available so far.

Doppler shifts, $v_i$, of the absorption and the emission components
of the P~Cygni profiles were determined separately
(Table~\ref{11044_p-cygni_my}). As the emission components are located
rather close to the rest wavelength and, except for lines from
\ion{Ne}{x}, are redshifted, and the absorption tails are, in
contrast, strongly blueshifted, they provide only upper/lower limits
of the velocity values.

\subsection{Line spectroscopy: Best-fit results} \label{sec:compilation}

\begin{figure*}\centering
\includegraphics[width=17cm]{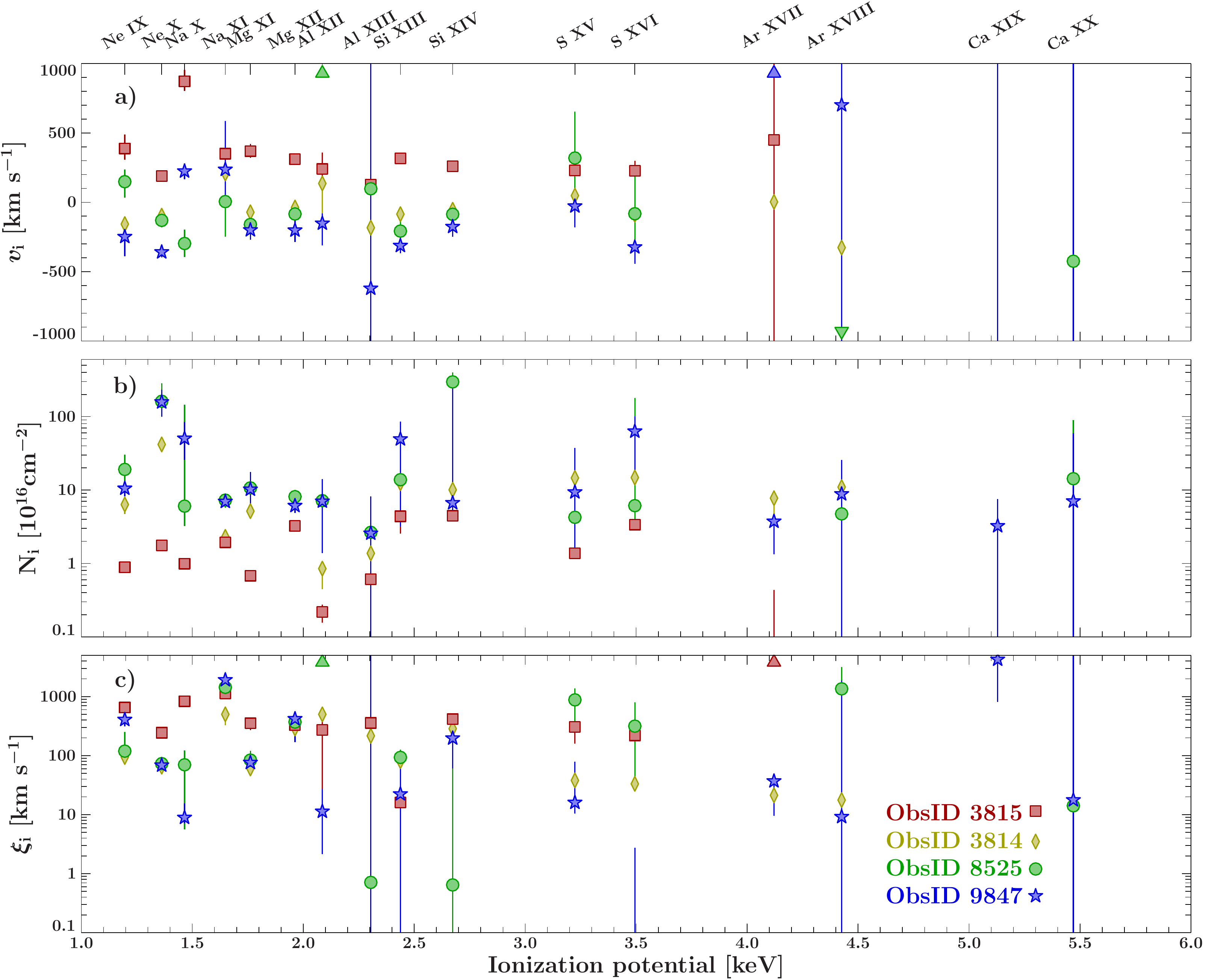}
  \caption{Fit parameters from Table~\ref{abslines} for line
    series of H-like and He-like transitions in ObsID~3815 (squares),
    ObsID~8525 (circles), ObsID~9847 (stars) and also
    ObsID~3814 (diamonds,  paper~I): \emph{top}: velocity
    shifts, \emph{middle:} column densities, 
    \emph{bottom:} thermal broadening. Colors correspond to ObsIDs as
    indicated. For completeness, values far above or below the
    displayed range are marked by triangles close to upper/lower
    $x$-axes.  
    \label{compilation_all}}
\end{figure*}

Figure~\ref{compilation_all} shows the parameters of individual
line series summarized in Table~\ref{abslines}: Doppler velocities
$v_i$, column densities $N_i$ and thermal broadening $\xi_i$ for
ObsIDs~8525,~9847, and~3815 ($\phi_\mathrm{orb}\sim0.0$--0.2 and ~0.75).
Parameters obtained in paper~I from ObsID~3814,
$\phi_\mathrm{orb}\sim0.95$, are
shown for comparison. Due to the different fitting approach, the
parameters of ObsID~11044 ($\phi_\mathrm{orb}\sim0.5$) are excluded from this
comparison.

All series in ObsID~3815 ($\phi_\mathrm{orb}\sim0.75$) are consistently
redshifted, with values falling in the interval between 100 and
$400\,\mathrm{km}\,\mathrm{s}^{-1}$. The exception is the \ion{Na}{x}
line, which is Doppler shifted by
$\sim$$870\,\mathrm{km}\,\mathrm{s}^{-1}$. The column densities of all
line series (Fig.~\ref{compilation_all}) are in the interval
$(0$--$5)\times10^{16}\,\mathrm{cm}^{-2}$ with a thermal broadening of
$\xi \lesssim 500\,\mathrm{km}\,\mathrm{s}^{-1}$, suggesting the lines
originate from the same region. 

Because of the short exposure of ObsIDs~8525 and 9847, many fit
parameters show large uncertainties or cannot be constrained. The
velocities in ObsIDs~8525 and 3814 ($\phi_\mathrm{orb}\sim0.0$) are much lower
compared to ObsID~3815, in fact, they are consistent with zero, while
the ones from ObsID~9847 ($\phi_\mathrm{orb}\sim0.2$) show mostly blueshifted
values. Overall, the velocities span the range $\pm
500\,\mathrm{km}\,\mathrm{s}^{-1}$.

The column densities of ObsIDs~8525, 9847, and 3814 are much higher
than those of ObsID~3815, supporting the idea of the denser focused
wind around $\phi_\mathrm{orb}\sim0.0$. We do not see any particular
trend in the thermal broadening measured in these observations,
although with $\lesssim$$300\,\mathrm{km}\,\mathrm{s}^{-1}$ it seems
to be generally lower than in ObsID~3815 (even though the broadening
shows a rather large scatter). For a plasma with a temperature of
$T\sim10^{6}$\,K, thermal velocities are on order of
$\sim10\,\mathrm{km}\,\mathrm{s}^{-1}$. The higher observed values
therefore suggest that most of the broadening is due to
microturbulence.

Closer investigation shows that for lines with lower signal-to-noise a
higher column appears to imply smaller thermal broadening. Such
behavior could explain the outlier lines. For example, in the case of
\ion{Si}{xiv} in ObsID~8525 assuming a typical value of
$\xi\sim$$300\,\mathrm{km}\,\mathrm{s}^{-1}$ would reduce the column
density by a factor of $\sim$10. For other lines this effect will be
smaller.

\section{The X-ray Orbital Variability of Cyg~X-1}\label{sec:variability}

Having described the properties of the individual observations, we now
turn to looking at them in the context of the different lines of sight
onto the black hole in order to understand better how the properties
of the wind depend on the orbital phase. We first study the variation
of the column density obtained from continuum or edge modeling with
orbital phase and then investigate the variation of the ionizing
absorber by looking at column densities and velocity shifts of
individual line series.

\begin{figure}\centering
\resizebox{\hsize}{!}{\includegraphics{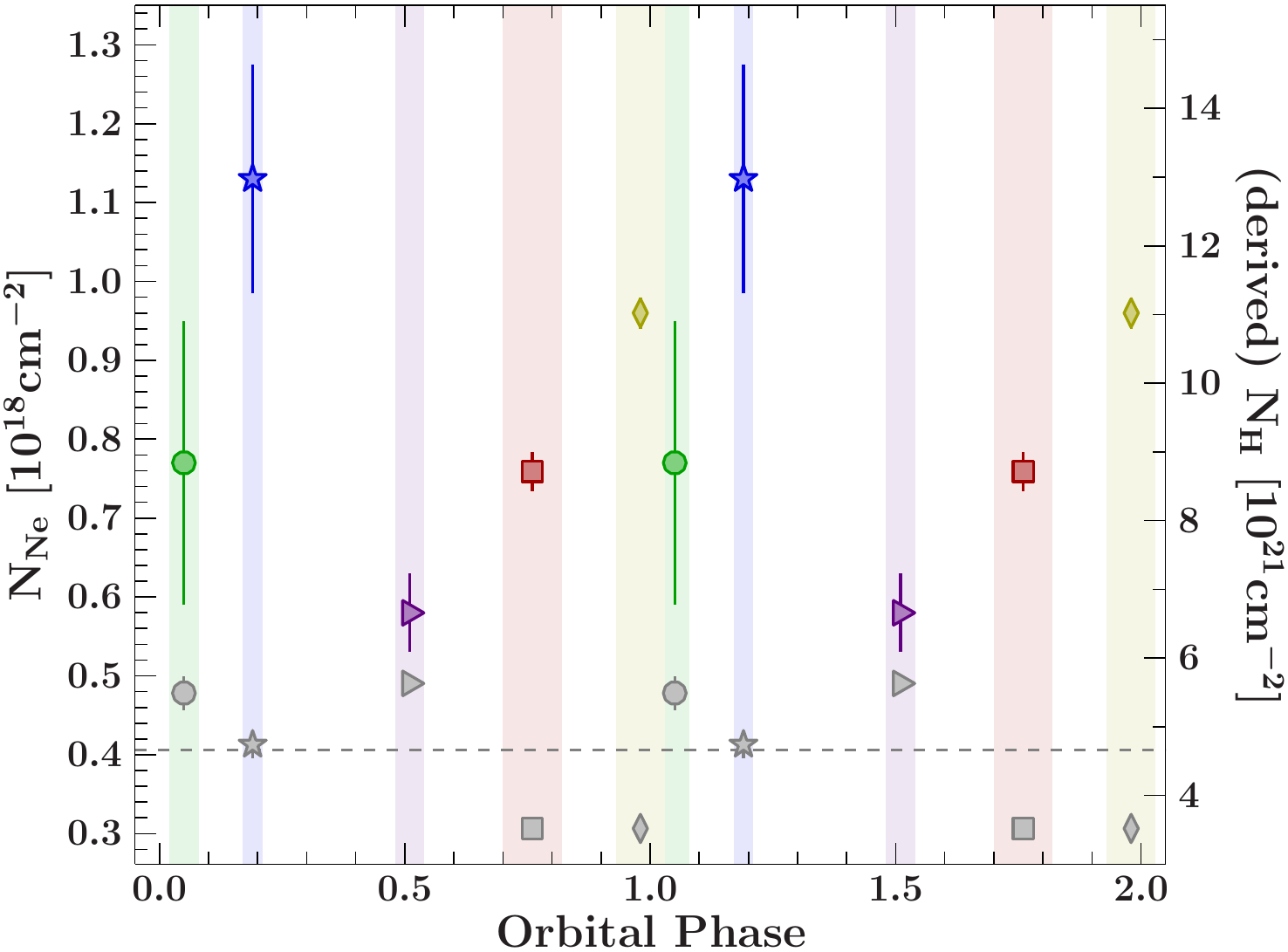}}
\caption{Orbital phase variation of fitted columns, repeated twice for
  clarity. Grey data points show the equivalent hydrogen column
  density, $N_\mathrm{H}$, from continuum fitting the \textit{Chandra}
  spectra (right hand $y$-axis). As discussed in the text, the data
  points are heavily influenced by systematics (the error bars shown
  are statistical only). The colored data points show the variation of
  the Ne neutral column from Ne-edge fitting (left $y$-axis) and
  corresponding $N_\mathrm{H}$ (right $y$-axis) derived assuming a
  Ne:H abundance of 1:11481 as per \citet{wilms00}. The phases of
  individual \textsl{Chandra} observations are denoted with colored
  regions. The dashed lines represent the total equivalent column
  density of the interstellar medium and of Ne
  \citep{Xiang11a}.}\label{neutral_orbital_variation}
\end{figure}

\begin{figure*}
\includegraphics[width=17cm]{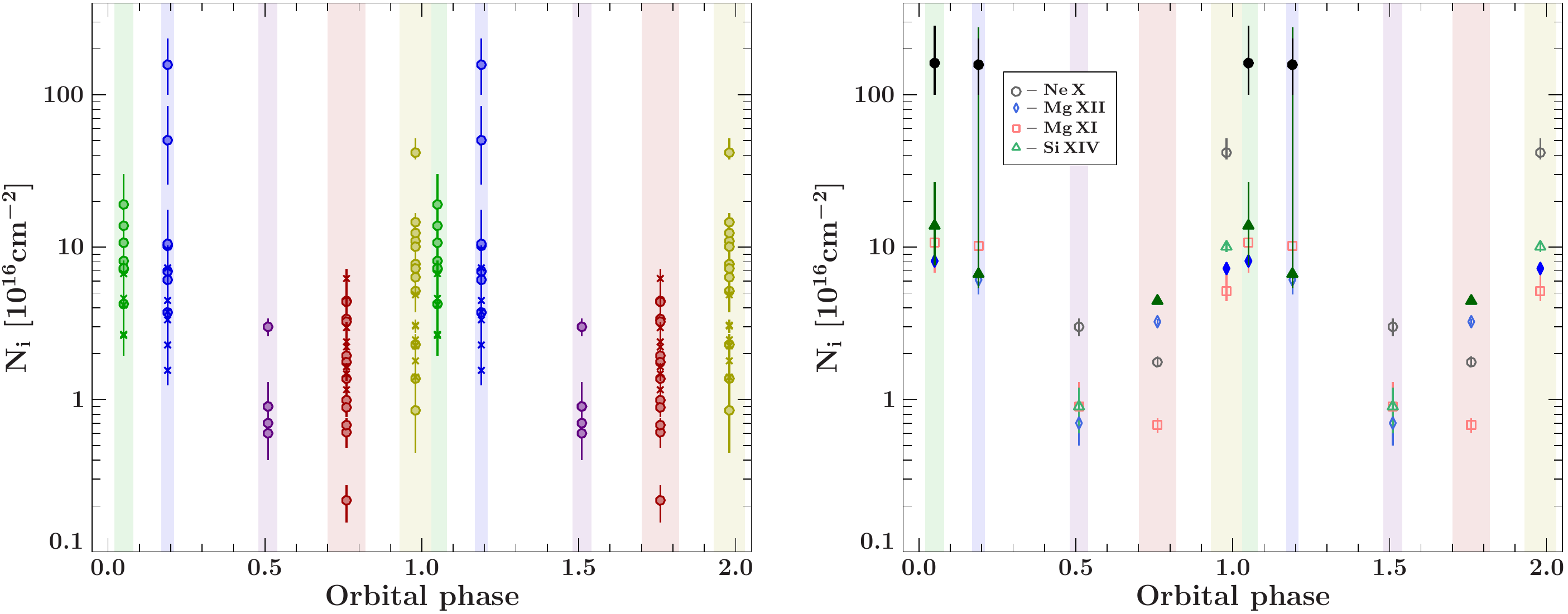}
\caption{Measured column densities, $N_i$, of individual elements as a
  function of orbital phase. \emph{Left}: H- and He-like line series
  (circles) and Fe lines (crosses). Only measurements with relative
  uncertainties $<$75\% are included. \emph{Right}: Column densities
  of the four most prominent ions, \ion{Ne}{x} (circles),
  \ion{Mg}{xii} (diamonds),
  \ion{Mg}{xi} (squares), and \ion{Si}{xiv} (triangles). Filled
  datapoints correspond to symmetric lines (see Fig.~\ref{asymmetry}
  and related discussions).} 
\label{variations_nh}
\end{figure*}

\subsection{Column density variation of the neutral absorber}\label{sec:neutnh}

The equivalent hydrogen column density $N_\mathrm{H}$ found from
continuum fitting is in principle a tracer for the material in the
interstellar medium along the line of sight to the X-ray source and
for the moderately ionized material in its vicinity. The reason is
that 0.5--10\,keV X-rays are mainly absorbed by K-shell electrons. As
the cross section of the K-shell is only mildly dependent on
ionization stage, absorption in moderately ionized material can
usually also be described reasonably well with cross sections for
neutral atoms, at least at the level that is usually used when
modeling broad band continua. $N_\mathrm{H}$ is thus a measure for
both the ISM foreground absorption and the column of mildly ionized
material in the stellar wind. Earlier monitoring, e.g., with RXTE,
showed a modulation of $N_\mathrm{H}$ with orbital phase
\citep{Grinberg_2015a,hanke11a,wilms06,kitamoto00,holt76}, although we
note that some of the modulation seen there is also caused by dips,
which were not removed in these analyses.

The values obtained from the \textit{Chandra} continuum fitting for
the $N_\mathrm{H}$ (Fig.~\ref{neutral_orbital_variation}, grey data
points) show a systematic offset compared to other determinations of
the column, with lowest values of
$\left(3.52^{+0.05}_{-0.03}\right)\times10^{21}\,\mathrm{cm}^{-2}$ at
$\phi_\mathrm{orb}\sim0.75$ and
$(3.52\pm0.04)\,\times10^{21}\,\mathrm{cm}^{-2}$ (paper~I) for
$\phi_\mathrm{orb}\sim0.95$. These values are smaller than the total
column density of the interstellar medium along the line of sight to
Cyg\,X-1, which dust scattering measurements determine to
$N_\mathrm{H_{tot}}\sim4.6\,\times\,10^{21}\,\mathrm{cm}^{-2}$
\citep[Fig.~\ref{neutral_orbital_variation}, dashed
  line;][]{Xiang11a}. This column sets the lower limit to the total
$N_\mathrm{H}$ to Cyg~X-1, which is the sum of the intrinsic
absorption in the system and the absorption in the interstellar
medium. Lower $N_\mathrm{H}$ values, as those seen here, thus indicate
a systematic error. This is not unlikely, when considering that the
lowest $N_\mathrm{H}$ values originate from an observation performed
in continuous clocking mode, where the continuum has to be modeled
with additional broad Gaussians (Table~\ref{cont_3815}). A further
systematic is the inability to constrain the continuum with
\textit{Chandra} in these absorption line dominated data. Our
experience shows that fits to simultaneous \textit{Chandra} and
\textit{RXTE}-PCA data indeed have significantly higher
$N_\mathrm{H}$. For example, in paper~I we found for observation 3814
that $N_\mathrm{H}$ increased from
$3.52\pm0.04\,10^{21}\,\mathrm{cm}^{-2}$ in the \textit{Chandra}-only
fits to $N_\mathrm{H}=5.4\pm0.4\,\times10^{21}\,\mathrm{cm}^{-2}$ when
including the PCA data, consistent with previous measurements
\citep{miller02,schulz02} and also with the $N_\mathrm{H}$ obtained
from continuum fitting to observation 8525, which has a similar
orbital phase. We therefore conclude that $N_\mathrm{H}$ values
obtained from continuum modeling the \textit{Chandra} data alone
suffer a large systematic uncertainty in these line dominated data,
with a clear systematic bias towards smaller $N_\mathrm{H}$. 

An independent, and likely better, measure for the absorbing column is
possible by direct measurement of the optical depth at absorption
edges. This approach is less dependent on systematics of the continuum
modeling than fitting $N_\mathrm{H}$, especially considering the
systematics induced by the brightness of the source. In our data, Neon
is the only element for which we can measure the column density
directly from the neutral edge in all of our observations
(Fig.~\ref{neutral_orbital_variation}, colored data points). The
observed variation suggests that part of the neutral $N_\mathrm{Ne}$
is intrinsic to the source and varies with orbital phase, with a
minimum at $\phi_\mathrm{orb}\sim0.5$, consistent with earlier studies
of the $N_\mathrm{H}$ variation in the system
\citep[e.g.,][]{Grinberg_2015a,wen99,kitamoto00}. We note, however,
that a potential systematic uncertainty is that only ObsID~11044
($\phi_\mathrm{orb}\sim0.5$) does not show any significant absorption
lines in the range of the Ne edge, while for the other ObsIDs this
region is populated by many strong lines from ionized Fe. It therefore
cannot be excluded that some of the variation is due to a
contamination of the Ne edge by these lines. A cross check in which
the continuum around the Ne edge is modeled locally reveals columns
that are systematically higher by a factor of $\sim$2, such that at
least part of the modulation could be due to the Fe lines. Local and
continuum modeling give similar results only for ObsID~3815
($\phi_\mathrm{orb}\sim0.75$). In addition to Ne, the good quality of
ObsID~11044 ($\phi_\mathrm{orb}\sim0.5$) allows us to leave the Fe
column a free parameter in the continuum fitting, giving
$N_\mathrm{Fe}=(0.15\pm0.02)\times10^{18}\,\mathrm{cm}^{-2}$ for this
observation. This value is consistent with $N_\mathrm{Fe}$ determined
from the local modeling of the edge and with the expected ISM Fe
contribution using the ISM abundances of \citet{wilms00}.

\subsection{Orbital variation of the column of the ionized absorber}
We now turn to the variation of the ionized absorber with phase.
Figure~\ref{variations_nh} illustrates the variation of the column
densities $N_i$ of highly ionized elements with orbital phase. The
parameter sample was reduced compared to the complete set of species
present in the spectra by excluding all columns with uncertainties
exceeding 75\% of their values. For ObsIDs~3814 and 3815
($\phi_\mathrm{orb}\sim0.95$ and 0.76) only one $N_i$ value per
observation, from \ion{S}{xvi} and \ion{Ar}{xvii}, respectively, was
lost. For ObsIDs~8525 and 9847 ($\phi_\mathrm{orb}\sim0.05$ and 0.2,
$\sim$4.4\,ks exposure) we discarded roughly half of the $N_i$ values
(ObsID~8525, $\phi_\mathrm{orb}\sim0.05$: H-like series of Ne, Al, Si,
S, Ar and Ca, and He-like Na; ObsID~9847, $\phi_\mathrm{orb}\sim0.2$:
both H- and He-like series of Al, Si, S, Ca, and H-like Ar).
Figure~\ref{compilation_all}b shows that these $N_i$ values are, to
within their uncertainties, consistent with the more tightly
constrained values. \ion{Ne}{x} and \ion{Na}{x} of ObsID~9847
($\phi_\mathrm{orb}\sim0.2$) appear as outliers, although compared to
other columns determined in this observation they are relatively well
constrained. In ObsID~11044 ($\phi_\mathrm{orb}\sim0.5$), where the
strong lines show P~Cygni profiles, the $N_i$ shown in
Fig.~\ref{variations_nh} were determined as described in
Sect.~\ref{sec:11044_ion_gas} for the absorption parts of
\ion{Ne}{x}$\alpha$, \ion{Mg}{xii}, \ion{Mg}{xi}, and \ion{Si}{xiv}.

There are four ions that are strong in all five observations and so
can be tracked over the whole orbit: \ion{Ne}{x}, \ion{Mg}{xii},
\ion{Mg}{xi}, and \ion{Si}{xiv} (Fig.~\ref{variations_nh}, right). On
closer investigation the lines appear asymmetric even in cases where
they do not show clear P~Cygni profiles. As asymmetry can affect the
column density measurements, we considered only those lines which do
not show any asymmetry according to Sect.~\ref{subsec:line-profiles}
(filled circles in Fig.~\ref{variations_nh}). Similarly to the cases
discussed above, $N_i$ is expected to reach its maximum at
$\phi_\mathrm{orb}\sim0.0$ and its minimum at $\phi_\mathrm{orb}\sim0.5$. The modulation
is clearly visible in Fig.~\ref{variations_nh}.

\subsection{The location of the absorbers}

\begin{figure*}\centering
\includegraphics[width=17cm]{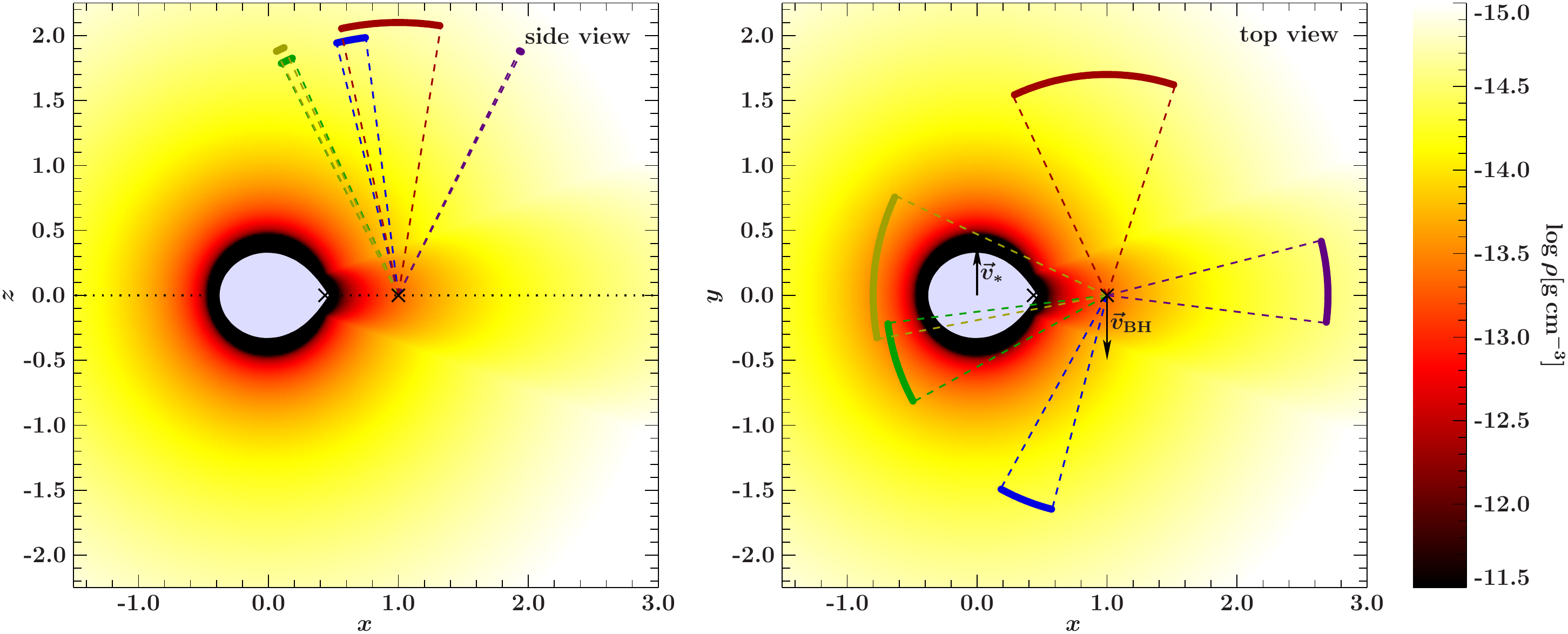}
\caption{Stellar wind density assuming the focused wind model of
  \citet{giesbolton86b} and lines of sight towards the black hole.
  \emph{Left}: Side view onto the orbital plane (dashed).
  \emph{Right}: Lines of sight projected onto the orbital plane. The
  black hole (cross) and its donor move clockwise around the
  center of mass (cross).}\label{fig:density}
\end{figure*}

\begin{figure}
\resizebox{\hsize}{!}{\includegraphics{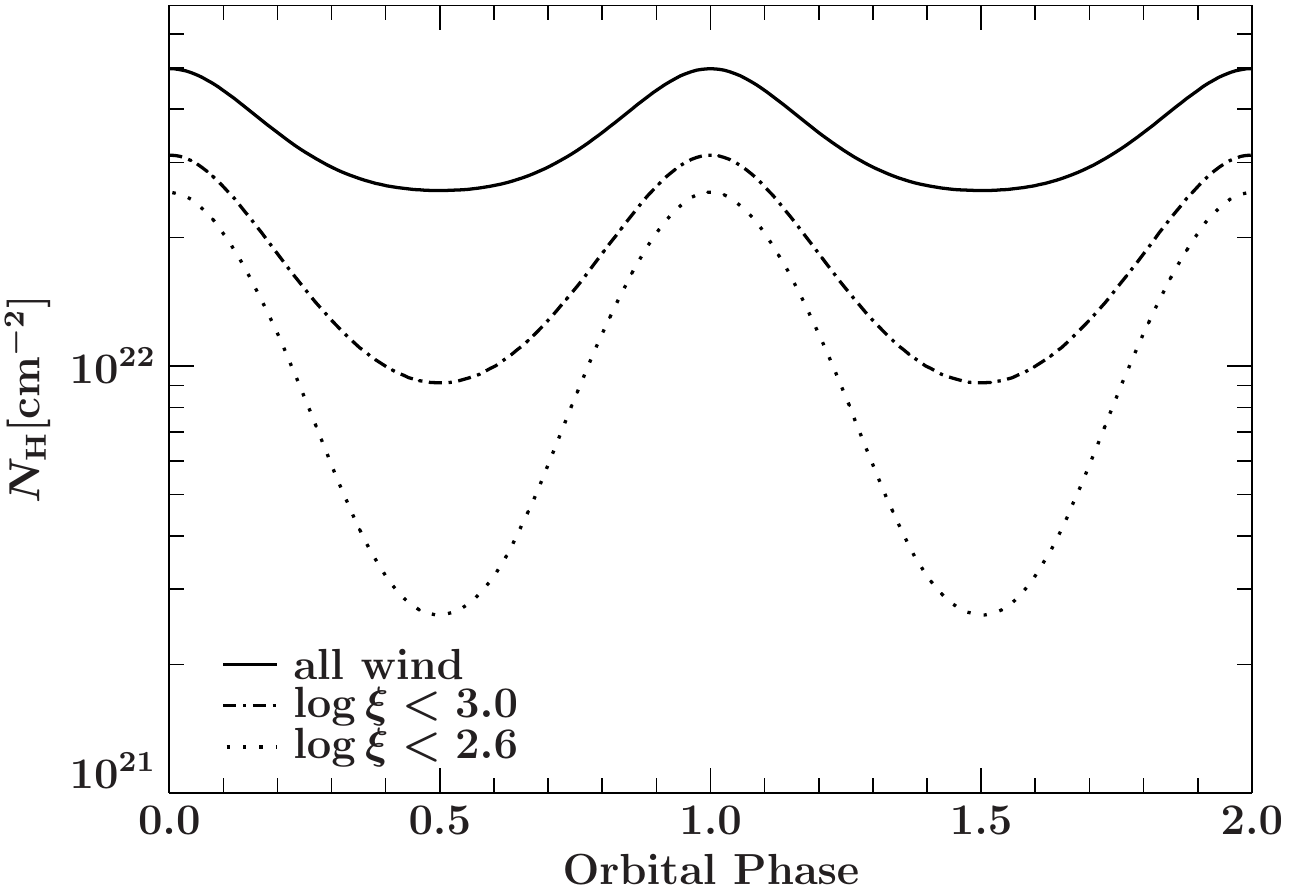}}
\caption{Phase variation of the column density according to the
  focused wind model of \citet{giesbolton86b}. Shown is the column of
  the whole wind (solid line), and of those less ionized wind regions
  where the ionization parameter $\log\xi<3.0$ (dash dotted line) and
  $\log\xi<2.6$ (dotted line), respectively. }\label{fig:theonh}
\end{figure}

The previous sections showed that both, the moderately ionized
material tracked by Ne-edge fitting and the highly-ionized material
responsible for the H- and He-like species shows orbital modulation,
i.e., the material is local to the X-ray binary. In this Section we
consider the relationship between both absorbers. First, as also
pointed out by \citet{marshall01}, note that in most observations the
He-like line series show lower columns than their H-like peers, $N_{i,
  \mathrm{He}}\,\leq\,N_{i, \mathrm{H}}$ (Fig.~\ref{compilation_all}),
which indicates that most of the material traced by the H- and He-like
lines is fully ionized. The best example in this respect is
ObsID~3815, where for all observed line series the column of the
H-like ions is larger than the column from the He-like ions. In this
ObsID, $\phi_\mathrm{orb}\sim0.75$, i.e., the line of sight is almost
perpendicular to the binary axis. For the other observations probing
the denser part of the wind, the relation between $N_{i, \mathrm{He}}$
and $N_{i, \mathrm{H}}$ is not as clear as for ObsID~3815, but $N_{i,
  \mathrm{He}}\leq N_{i,\mathrm{H}}$ is fulfilled for most ions.

To compare the measured variation of the columns with theoretical
expectations we use the focused wind model of \citet{giesbolton86b} as
a toy model. This model consists of a CAK-model with a longitudinal
variation of the wind parameters in a cone $\pm 20^\circ$ degrees from
the line between the donor and the black hole. Outside of that region
we use the wind parameters at $\theta=20^\circ$, where $\theta$ is the
angle in the orbital plane of the binary, measured from the line
between the donor and the black hole. Figure~\ref{fig:density} shows
the density structure of the focused wind as well as the line of sight
of our observations projected onto the orbital plane and in a side
view. The model has been shown to be a good overall representation of
the stellar wind in the HDE 226868/Cyg X-1 system, even though the
detailed parameters of the wind are still debated \citep[][and
references therein]{gies03,gies08,vrtilek:08a}. We emphasize that much
of this discussion relates to the wind properties in the so-called
``shadow wind'', i.e., the region of the star opposite to the black
hole where X-rays are not presumed to influence the wind properties
\citep{caballero09}. This region is not probed by the line of sights
studied here.

Figure~\ref{fig:theonh} (solid line) shows that in the model of
\citet{giesbolton86b} the total (neutral and ionized) column is around
$N_\mathrm{H}\sim4\times 10^{22}\,\mathrm{cm}^{-2}$, i.e., it is more
than a factor of 10 higher than the $N_\mathrm{H}$ measured from
fitting the continuum or the $N_\mathrm{H}$ values inferred from
fitting the Ne edge. The latter vary between $6.8\times
10^{21}\,\mathrm{cm}^{-2}$ and $1.3\times 10^{22}\,\mathrm{cm}^{-2}$
(Fig.~\ref{neutral_orbital_variation}). Due to the expected
high degree of ionization caused by the black hole and the UV
radiation from its donor, this difference between the measured
$N_\mathrm{H}$ and the model predicted total columns is not surprising
\citep[e.g.,][]{holt76,wen99}. Figure~\ref{fig:theonh} therefore also
shows the expected columns from regions where the ionization
parameter, $\log\xi$, is $<3$ and $<2.6$. Following
\citet{tarter:69a}, we define the ionization parameter as
$\xi=L/(nr^2)$, where $L$ is the source luminosity above the hydrogen
Lyman edge, $n$ is the absorbing particle density and $r$ is the
distance from the ionizing source. We will be using
$L=10^{37}\,\mathrm{erg}\,\mathrm{s}^{-1}$ throughout, a typical value
for our observations when taking into account the UV and $>$10\,keV
emission of the source (Table~\ref{observations}). Similar to
\citet{wen99} we also ignore the contribution of the UV photons from
the donor star, which systematically increases the ionization
parameter. Numbers quoted in the following are therefore to be
considered lower limits. Note that we use the symbol $\xi$ for both,
the ionization parameter and the thermal broadening parameter in line
series fitting. We do not anticipate confusion because they are used
in different contexts.

Figure~\ref{fig:theonh} indicates that the modulation of the
absorption found in the Ne edge and in line series
(Figs.~\ref{neutral_orbital_variation} and~\ref{variations_nh}) is in
qualitative agreement with the variation expected due to the
moderately ionized region of the wind, i.e., where $\log\xi<2.6$. This
is the outermost region of the wind (for $r\rightarrow \infty$,
$\log\xi\rightarrow 2.5$), i.e., consistent with the observational
findings most of the wind is expected to be ionized. With a variation
by a factor of $\sim$2, however, the total expected variation in the
model of \citet[][Fig.~\ref{fig:theonh}, solid line]{giesbolton86b} is
at the lower end of what is found in the observations, where we found
a variation of a factor of $\sim$2 for tracers of the neutral column
(Fig.~\ref{neutral_orbital_variation}, colored data points) to a
factor of $\sim$10 for the highly ionized species
(Fig.~\ref{variations_nh}). In addition, comparing the expected
variation at low $\log\xi$ with the total column, based on
Fig.~\ref{fig:theonh} we would expect ionized columns that are only a
factor of a few larger than the neutral column, while for most ions
the difference in the data is higher. It is unlikely that all of this
discrepancy is due to the simplifications of the wind model of
\citet{giesbolton86b} alone. We note, however, that we used the
interstellar abundances to convert the overall continuum
$N_\mathrm{H}$ values to columns for the individual ions. This is
probably not correct, since an evolved star such as HDE 226868 is
likely to show a different abundance pattern. Assuming that the most
important metals are overabundant with respect to the interstellar
medium, the difference could be explained. Overabundances in Cyg\,X-1
were reported previously. Modeling the emission K$\alpha$ line,
\citet{duro11} and \citet{fabian12} determined the Fe abundance in the
accretion disk to be 1.2--$1.6A_{\mathrm{Fe}_\odot}$. Further analysis
gives much higher values of
$A_\mathrm{Fe}\sim3$--$6A_{\mathrm{Fe}_\odot}$
\citep{duro13}. Reflection modeling also yields an iron
overabundance of 1.9--2.6 \citep{tomsick14}. Independent of X-ray
measurements, the analysis of optical and UV spectra of HDE~226868
also yields a nitrogen overabundance of five times solar
\citep{caballero09}. This result also suggests the possibility of the
other elements being overabundant, which is not so surprising, given
that the system consists of an evolved star and has experienced a
supernova.

We therefore conclude that a large fraction of the observed medium is
ionized and that it is very likely that the observed wind volume is
enriched in metals. Since the medium is ionized, it is located fairly
close to the black hole. Depending on the line of sight, the distance
from the black hole within which $\log\xi>2.6$ is between $0.5d$ and
$3.6d$, where $d$ is the distance between the donor and the black
hole. To constrain the properties of the absorber further, we
therefore need to turn to another observable, the Doppler shift of the
observed lines.

\begin{figure}
\resizebox{\hsize}{!}{\includegraphics{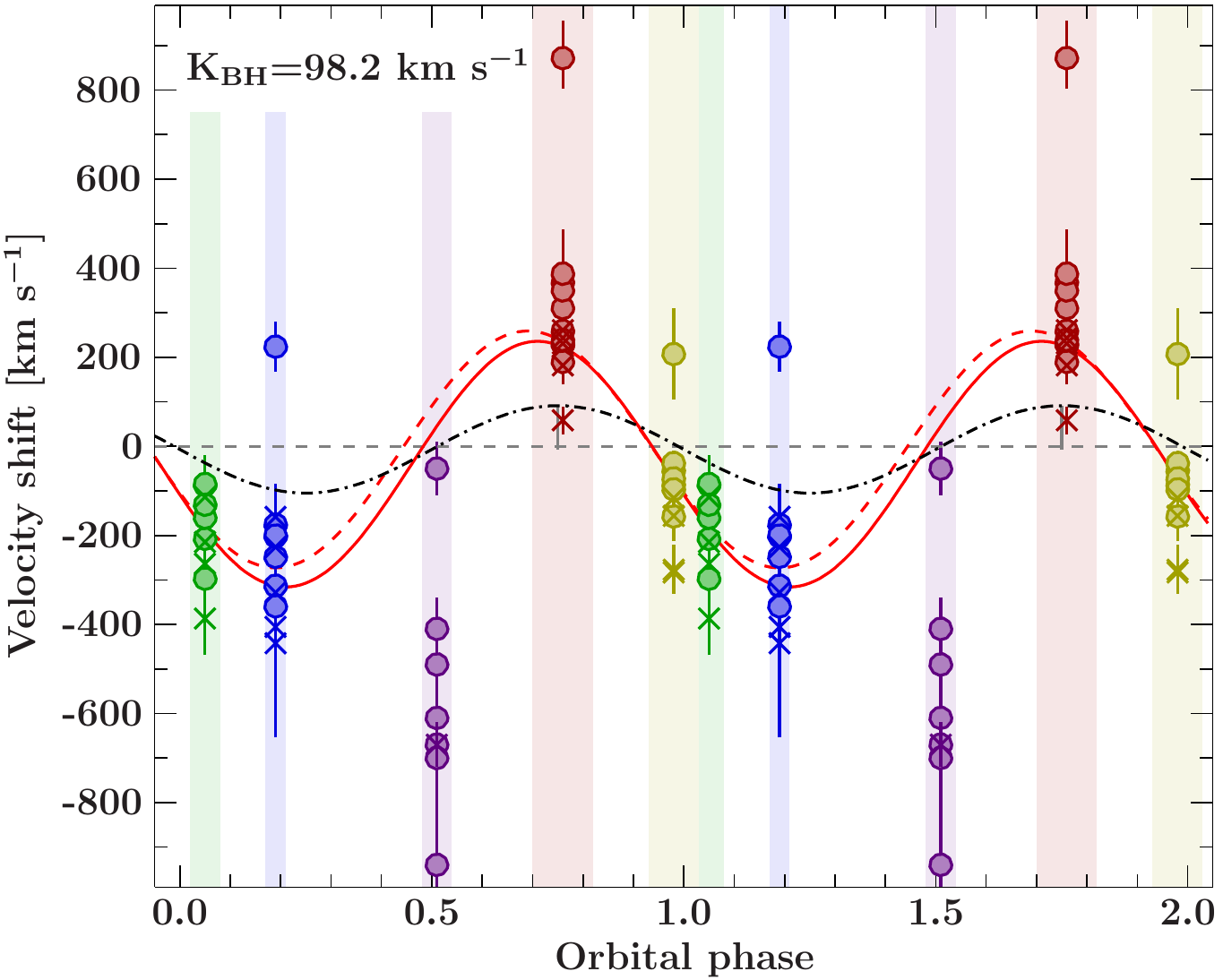}} 
\resizebox{\hsize}{!}{\includegraphics{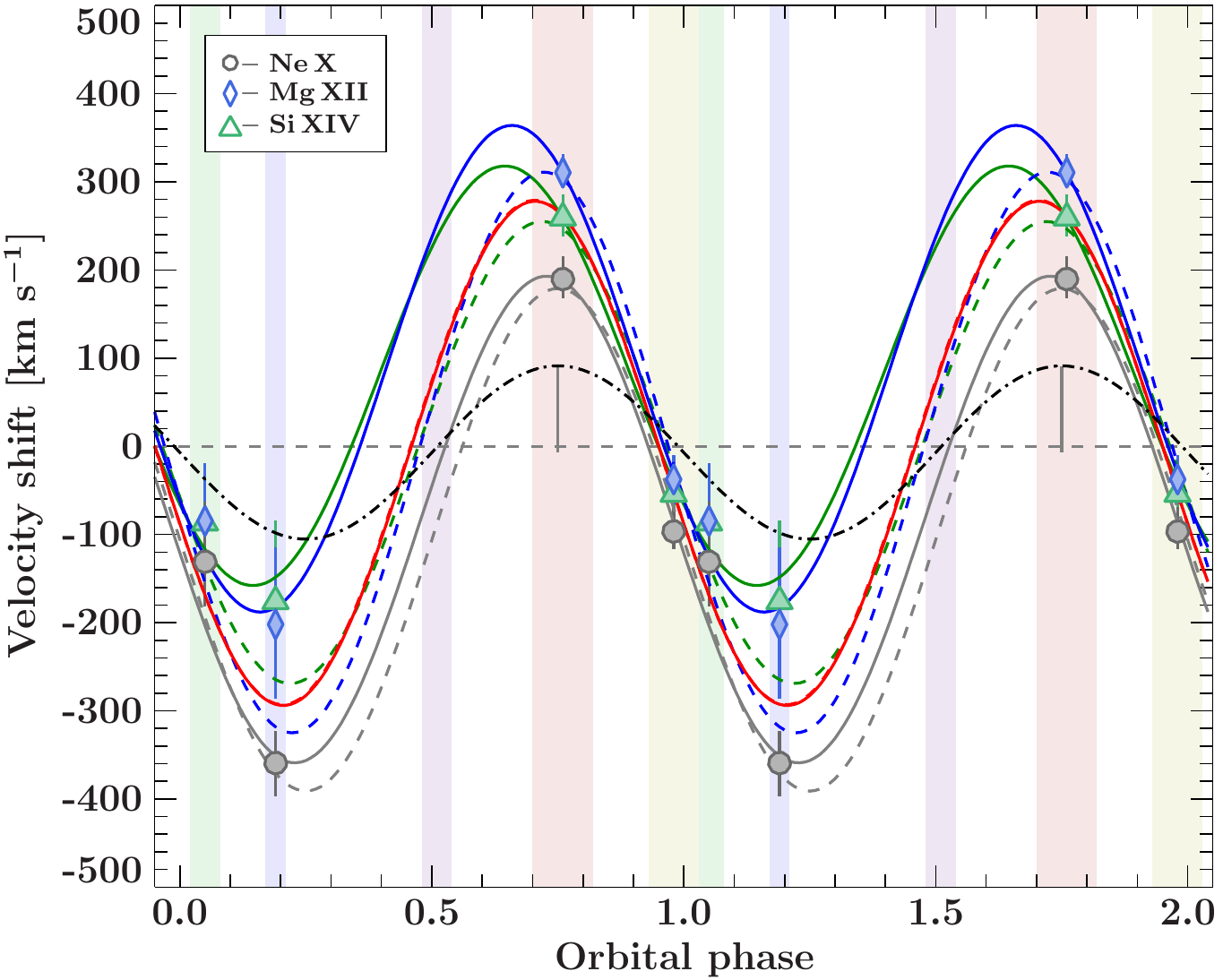}}
\caption{\emph{Top}: Velocity shifts of individual line series
  (circles: H-like and He-like ions, crosses: Fe line series) and the
  radial velocity of the black hole (dash-dotted line, semi-amplitude
  of $\sim$$98\,\mathrm{km}\,\mathrm{s}^{-1}$). The red sine curve
  shows the best fit involving all series shown
  ($K_{W}\sim276\,\mathrm{km}\,\mathrm{s}^{-1}$,
  $v_0\sim-40\,\mathrm{km}\,\mathrm{s}^{-1}$, and
  $\Delta\phi\sim0.46$), while the dashed line shows the ``second best
  fit'' where the offset was fixed to the systemic velocity of the
  black hole (Table~\ref{amplitudes}). \emph{Bottom}: Velocity
  shifts for individual ions and the radial velocity of the black hole
  (dash-dotted line). The sine curves represent the best fits for each
  of the three given line series (Table~\ref{amplitudes}). Dashed
  lines show fits with the offset fixed to the systemic velocity of
  the black hole. The red sine curve is the best fit involving all
  three line series (Table~\ref{amplitudes}).}
 \label{vel_variations_diff_orb_phases}
\end{figure}

\begin{table}
  \caption{Sine function parameters describing the orbital modulation
    of the  Doppler shifts. The fits are to a model of the form
    $v(\phi)=v_0+K_\mathrm{W}\cdot\sin[2\pi(\phi-\Delta\phi)]$. Parameters
    without uncertainties were held fixed in the
    fit.} \label{amplitudes}
\centering
\begin{tabular}{lllll} 
\hline 
\hline
 $K_\mathrm{W}$ & $v_0$ & $\Delta\phi$ & $\chi^{2}/\mathrm{dof}$ & $\chi^{2}_\mathrm{red}$  \\ 
\ [$\mathrm{km}\,\mathrm{s}^{-1}$] & [$\mathrm{km}\,\mathrm{s}^{-1}$] & & & \\
\hline
\multicolumn{5}{l}{all lines included in Fig.~\ref{vel_variations_diff_orb_phases}:}\\
 $\bf276\pm11$ &  $\bf -40\pm14$ & $\bf0.073\pm0.002$ & $\bf428.28/34$   & $\bf12.59$ \\ 
 $242\pm10$ & $-7$       & $0$               & $539.01/36$   & $14.97$ \\
$\bf266\pm10$ & $\bf-7$       & $\bf0.070\pm0.002$   & $\bf433.93/35$  & $\bf12.39$ \\
 $297\pm11$ & $-84\pm8$  & $0$               & $442.79/35$  & $12.65$  \\
\hline
\multicolumn{5}{l}{fit to \ion{Ne}{x}, \ion{Mg}{xii}, and \ion{Si}{xiv}:}\\
$\bf286^{+25}_{-24}$ & $\bf-8\pm36$ & $\bf0.072\pm0.004$ & $\bf30.87/9$   & $\bf3.43$ \\ 
$271\pm19$ & $-7$       & $0$               & $63.45/11$   & $5.77$ \\
$\bf286\pm20$ & $\bf-7$ & $\bf0.072\pm0.002$ & $\bf30.9/10$  & $\bf3.09$ \\
$313\pm24$ & $-62\pm18$  & $0$               & $39.07/10$  & $3.91$  \\
\hline
\multicolumn{5}{l}{fit to \ion{Ne}{x}:}\\
$\bf276\pm37$& $\bf-83\pm48$  & $\bf0.076\pm0.006$ & $\bf1.72/1$   & $\bf1.72$ \\ 
$236\pm33$ & $-7$       & $0$             & $41.24/3$  & $13.75$ \\
$267\pm34$ & $-7$       & $0.067\pm0.003$ & $8.49/2$    & $4.24$ \\
$\bf285\pm36$ & $\bf-106\pm26$ & $\bf0$             & $\bf2.67/2$ & $\bf1.33$  \\
\hline
\multicolumn{5}{l}{fit to \ion{Mg}{xii}:}\\
$\bf276^{+47}_{-38}$ & $\bf88\pm91$ & $\bf0.07\pm0.01$ & $\bf0.48/1$ & $\bf0.48$ \\
$309\pm30$   & $-7$     & $0$                & $7.67/3$  & $2.56$ \\
$\bf318\pm32$   & $\bf-7$     & $\bf0.075\pm0.003$    & $\bf3.46/2$ & $\bf1.73$ \\
$334\pm47$   & $-33\pm38$ & $0$                &           $6.39/2$ & $3.19$\\
\hline
\multicolumn{5}{l}{fit to \ion{Si}{xiv}:}\\
$\bf238^{+48}_{-41}$ & $\bf80\pm89$ & $\bf0.063\pm0.013$ & $\bf0.69/1$ & $\bf0.69$ \\
$256\pm38$   & $-7$     & $0$                & $6.79/3$  & $2.26$ \\
$\bf262\pm38$   & $\bf-7$     & $\bf0.075\pm0.004$    & $\bf3.31/2$ & $\bf1.65$ \\
$275\pm48$   & $-30\pm37$ &  $0$                & $5.74/2$ & $2.87$\\
\hline  
\hline 
\end{tabular}
\tablefoot{$K_\mathrm{W}$ -- amplitude of the sine curve; $v_0$ --
  offset of the sine curve, $\Delta\phi$ -- phase shift:
  $v_0=-7\,\mathrm{km}\,\mathrm{s}^{-1}$ and $\Delta\phi=0$ are fixed
  to the orbital motion of the black hole. Best fits are
  printed in \textbf{bold}.}
\end{table}

\begin{figure}
\resizebox{\hsize}{!}{\includegraphics{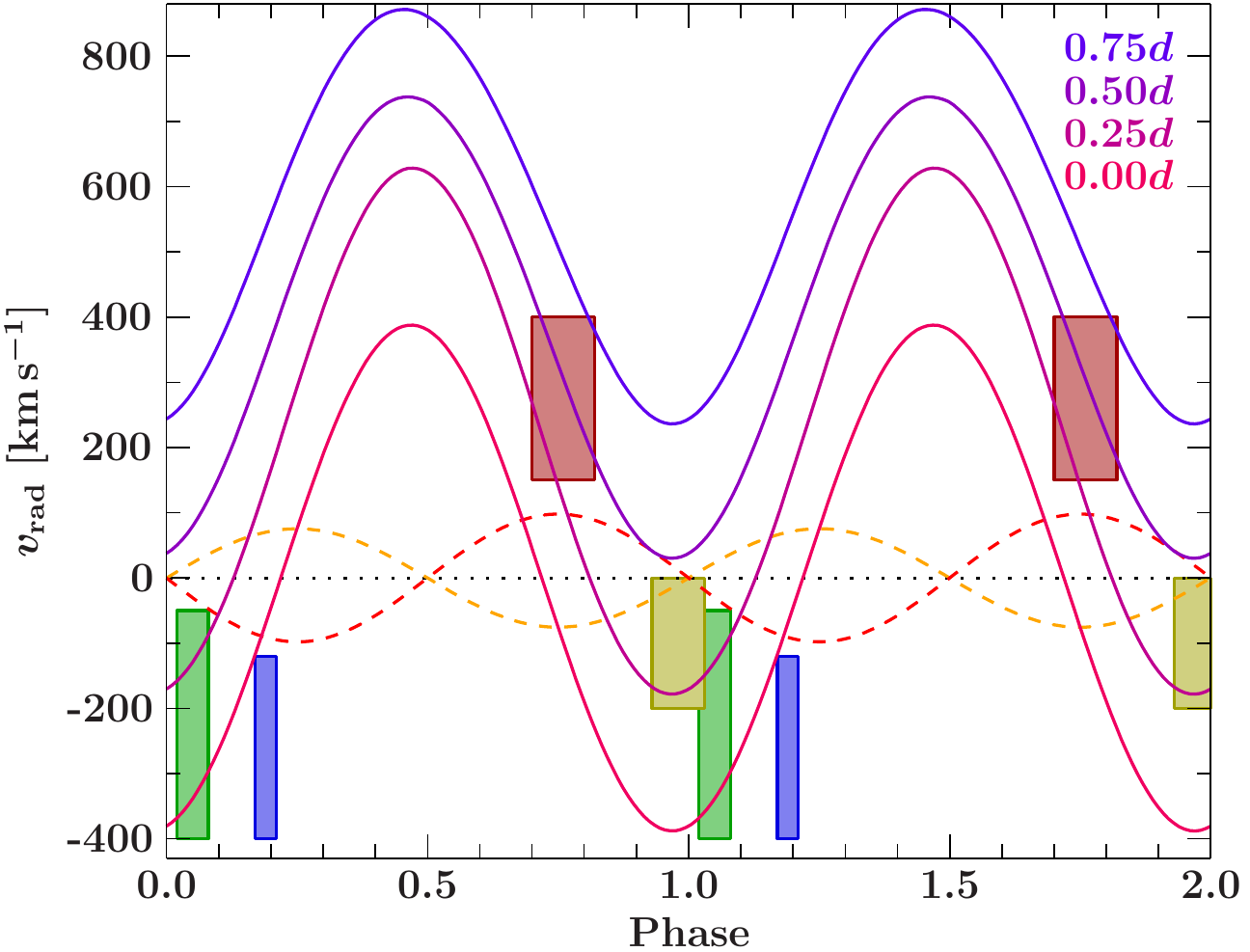}}
\caption{Radial velocities of the black hole (red dashed line) and of
  HDE 226868 (orange dashed line) and wind velocities projected onto
  the line of sight as a function of distance from the black hole.
  Distances are given in units of the distance between the black hole
  and its donor, $a$. Colored boxes indicate the range of the measured
  Doppler shifts from the \textit{Chandra} observations, excluding ObsID
  11044.}\label{fig:vrad}
\end{figure}

Figure~\ref{vel_variations_diff_orb_phases} shows that there is a
systematic sinusoidal modulation of the velocity shifts of individual
line series from the ionized absorber with orbital phase. Velocities
for ObsID~11044 at $\phi_\mathrm{orb}\sim 0.5$ are systematically
shifted to lower velocities because here only the absorbing part of
the P~Cygni profile was taken into account. We therefore ignore this
observation for the remainder of our discussion.

At a first glance, the line shifts appear to be approximately in phase
with the radial velocity of the black hole, which has a velocity
amplitude of $K_\mathrm{BH}\sim98\,\mathrm{km}\,\mathrm{s}^{-1}$
\citep{orosz11} and a systemic velocity of
$v_0=-7.0\pm0.5\,\mathrm{km}\,\mathrm{s}^{-1}$ \citep{gies03}. We
performed a set of fits to establish whether the wind follows the
black hole, and investigated several different scenarios
(Table~\ref{amplitudes}): we used all line series shown in
Fig.~\ref{vel_variations_diff_orb_phases} including velocities from
ObsID~11044, $\phi_\mathrm{orb}\sim0.5$, the lines of \ion{Ne}{x},
\ion{Mg}{xii}, \ion{Si}{xiv} together, and each of these three lines
individually. Apart from a fit in which all parameters of the sine
were left free, we also constrained the fits to be in phase with the
black hole and to have a different amplitude than the black hole.
Generally, models with varying phase shifts give better results than
those with the zero phase shifts (bold rows in
Table~\ref{amplitudes}). The velocities of different ions are clearly
not consistent with each other, indicating a complex ionization
structure.

We contrast these results in Fig.~\ref{fig:vrad} with the radial
velocities expected from the focused wind model of
\citet{giesbolton86b}. A direct fit of the ion velocities to this
model is not justified, the figure therefore indicates the typical
range for the Doppler shifts of our observations. The projected wind
velocities for the range of distances in which the highly ionized
lines originate according to our analysis of the columns is in
agreement with the measured shifts. Note, however, that the wind model
predicts a phase shift of $\Delta\phi=0.25$ between the black hole and
the wind, different from what is seen. The measurements during phase 0
are consistent with an origin $\lesssim 0.25d$ from the black hole (as
before, $d$ is the distance between the donor and the black hole),
while the data taken during phase 0.75 are consistent with a distance
$\lesssim 0.5d$. Given that this observation is the brightest of our
observations, it is not unexpected that the ionized region should move
away from the black hole. The overall redshift measurements therefore
agree with our conclusion that the ionized absorber must originate
close to the black hole, and possibly in the focused wind.

While the data from observations 3814, 3815, and 8525 could originate
in the focused wind, ObsID 9847 presents an interesting outlier.
During this observation the absorbing material appears to be moving
towards the observer with a blueshift that is much higher than that of
the black hole. As shown in Fig.~\ref{fig:vrad}, except for a region
very close to the black hole, only positive radial velocities
(=redshifts) are expected during this observation. To reconcile these
data with a focused wind would require terminal velocities of
$>4500\,\mathrm{km}\,\mathrm{s}^{-1}$. This value is much higher than
the $\sim$$2500\,\mathrm{km}\,\mathrm{s}^{-1}$ typically assumed for
the wind at the location of the black hole (this value decreases for
stream lines away from the line of symmetry of the system). As an
alternative explanation we note that during this observation we are
looking through the ``bow shock'' of the black hole \citep[e.g.,][and
references therein]{manousakis:11a,blondin95a}, i.e., a region where
the wind is strongly disturbed. It is not unlikely that in this region
material gets so disturbed that it obtains a significant non-radial
velocity, which at this phase would show up as a high blue shift.

\section{Advanced Line Profile Studies}\label{sec:line-profiles}
\subsection{P~Cygni Profiles and the Stellar Wind}\label{subsec:line-profiles}

\begin{figure*}\centering
\centering\includegraphics[width=17cm]{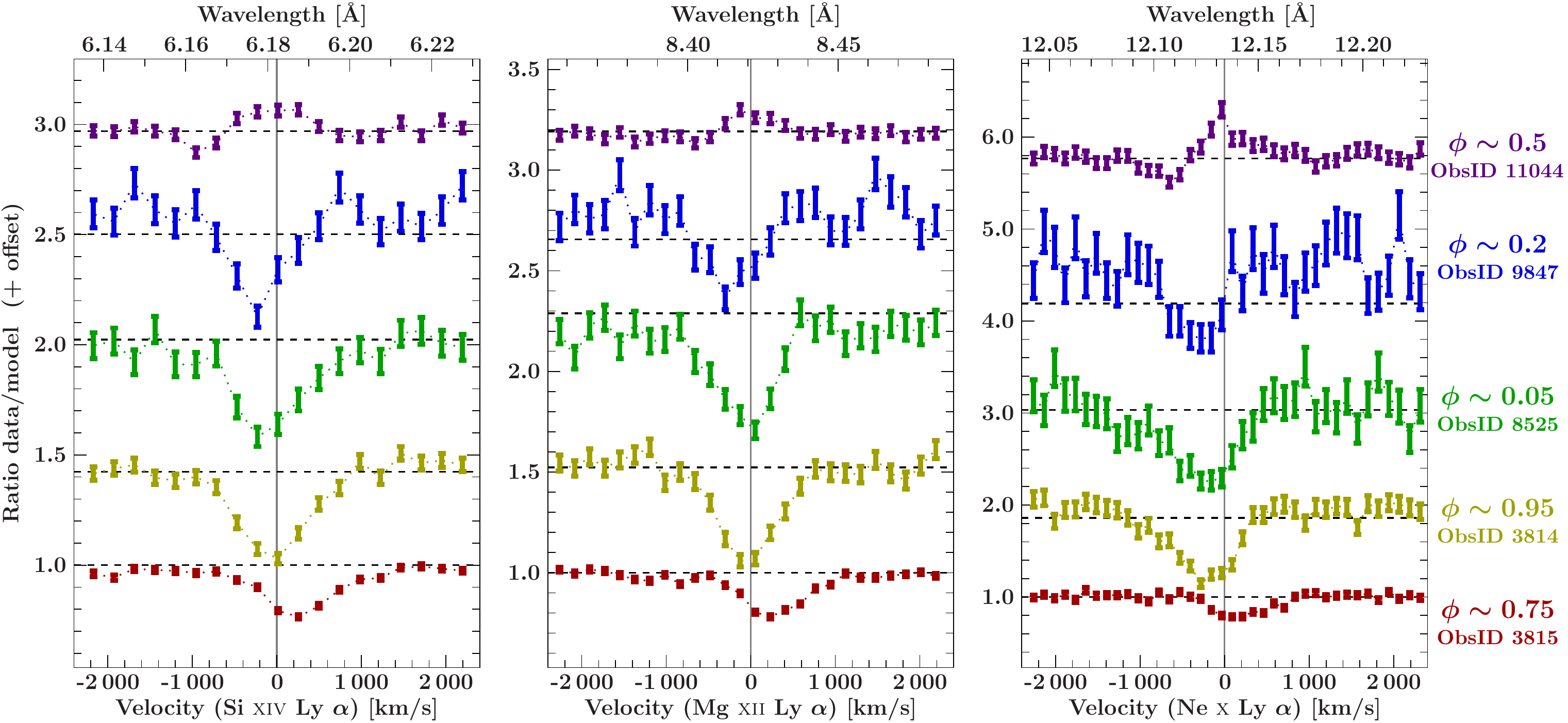}
\caption{Line profiles of the three most prominent absorption lines of
  \ion{Si}{xiv}, \ion{Mg}{xii}, and \ion{Ne}{x} are shown
  here as they vary over the orbital phase. Lines are displayed as the
  ratio of data and model, offset for visual clarity. For
  each line, the vertical grey lines are located at rest wavelengths
  corresponding to zero velocity shifts. See text  for explanation. 
  \label{all-line-profiles}}
\end{figure*} 

\begin{figure*}\centering
\centering\includegraphics[width=17cm]{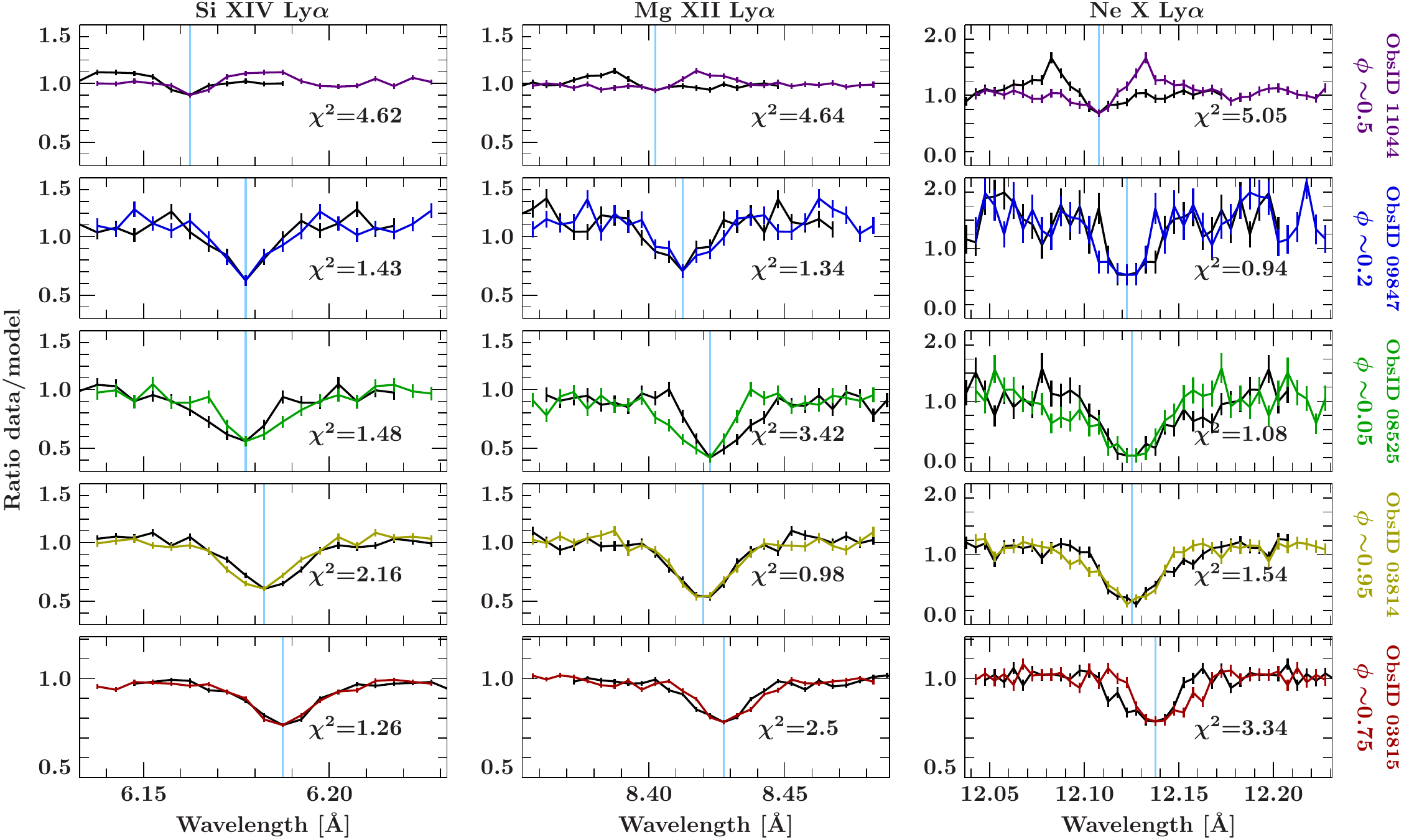}
\caption{The same line profiles as in Fig.~\ref{all-line-profiles},
  but this time investigating the symmetry of the profiles. The chosen
  absorption lines have clear P~Cygni profiles at phase
  $\phi_\mathrm{orb}\sim0.5$. Also other lines show, however, asymmetric
  profiles. The colored profiles show the original ones, while the
  black profiles are mirrored at the energy bin with the lowest
  relative flux value.    
  \label{asymmetry}}
\end{figure*} 

In the previous section we already alluded to the complexity of some
line profiles. We now discuss the behavior of the absorption lines
with the highest signal-to-noise ratio, i.e., the \ion{Si}{xiv},
\ion{Mg}{xii}, and \ion{Ne}{x} Ly $\alpha$ transitions
(Fig.~\ref{all-line-profiles}). We observe strong, pure absorption
lines with small blueshift \citep[no more than
$200\,\mathrm{km}\,\mathrm{s}^{-1}$,][and paper~I]{hanke08}, or no
shift at all between phases $\phi_\mathrm{orb}\sim0.0$ and $\sim0.2$. Lines at
phase $\phi_\mathrm{orb}\sim0.5$ and $\sim$0.75 show, however, very different
profiles. At phase $\phi_\mathrm{orb}\sim0.5$, we detected \emph{clear} P~Cygni
line profiles in the spectrum of Cyg\,X-1. Variable P~Cygni-like line
profiles have previously been seen, e.g., in Cir~X-1
\citep{schulz02a,schulz08a}. In Cyg\,X-1, P~Cygni lines have previously
been suggested by \citet{schulz02} from data from ObsID~107. This
15\,ks TE mode observation observation at phase $\phi_\mathrm{orb}=0.73$--0.76
was performed during a softer, higher flux state of the source. The
observed spectral features were mostly identified with Fe transitions
\ion{Fe}{xviii}--\ion{Fe}{xxiv}, and the authors suggested that
absorption and close emission lines could form P~Cygni profiles in
some cases. They argued, however, that the identification of the Fe
lines in the range between 6 and 16\,\AA\ is difficult, as many of
them are probably blended. Investigating ObsID 2415 at the same
orbital phase, \citet{miller05} did not find convincing evidence of
P~Cygni features. Our spectrum of the observation at $\phi_\mathrm{orb}\sim0.75$
(ObsID~3815) does not reveal any P~Cygni line profiles, but we note
that the lines appear asymmetric (see Fig.~\ref{asymmetry} and
discussion below).

Given the values from Table~\ref{11044_p-cygni_my}, the emission
component is much stronger than the absorption. It appears slightly
blueshifted, at very low velocities that are -- given their
uncertainty -- still consistent with zero. The blueshift of the weak
absorption tails is much higher. Line profiles at phase
$\phi_\mathrm{orb}\sim0.75$ are observed as pure absorption lines, but are much
weaker than at $\phi_\mathrm{orb}\sim0.0$ and redshifted by
100--$400\,\mathrm{km}\,\mathrm{s}^{-1}$. The high blueshift observed
in the P~Cygni profiles is a direct proof that a non-focused part of
the wind also exists on the X-ray irradiated side of the donor star,
and that it is accelerated to high velocities. If the different line
profiles at $\phi_\mathrm{orb}\sim0.0$ and $\sim0.5$ result from the same wind
structure, only seen at different viewing angles, then the weakness of
the absorption lines at $\phi_\mathrm{orb}\sim0.5$ confirms that the line of
sight at this phase probes less dense regions of the wind than at
$\phi_\mathrm{orb}\sim0.0$. Since no net emission is seen at $\phi_\mathrm{orb}\sim0.0$,
any emitting gas must be at a small projected velocity, where the
emission fills only part of the absorption trough in the continuum
radiation created along the line of sight. As this gas is also seen at
low velocity from $\phi_\mathrm{orb}\sim0.5$, its full space velocity must be
small. It could, e.g., be related to the slow base of the spherical
wind, or to the focused wind.

To find possible constraints on the wind in the system,
\citet{miller05} used the observed lines (\ion{Ne}{ix}, \ion{Ne}{x}
\ion{Si}{xiii}, and \ion{Si}{xiv}). Based on calculations by
\citet{kallman82} and using the spectral shape and luminosity
appropriate for their observations, they find that these absorption
features arise in a temperature region spanning over one order of
magnitude as $\log T=4.5$--6.0. This result already implies that even
by investigating the non-dip spectrum only, we probe a
multi-temperature region. Based on the ionization parameter and
simplifying geometry assumptions, \citeauthor{miller05} determined the
distance of the absorbing region from the ionizing source to
be $\sim3\times10^{11}-10^{12}\,\mathrm{cm}$.

Most of the line profiles shown in Fig.~\ref{all-line-profiles} give
an impression of asymmetry. To describe it quantitatively, we first
mirrored the measured line profile at the energy bin with the lowest
relative flux value of the data/model ratio (Fig.~\ref{asymmetry}).
The original and the flipped profiles were then treated as ``data''
and ``model'' and a $\chi^{2}$-type value was calculated to describe
the ``goodness'' of the overlap according to
\begin{equation}
\chi^{2}=\frac{1}{N}\sum\frac{(y_\mathrm{original}-y_\mathrm{flipped})^{2}}{\sigma_{y_\mathrm{original}}^{2}+\sigma_{y_\mathrm{flipped}}^{2}}
\label{chi-squared}
\end{equation} 
where $N$ is the total number of bins which were taken into account to
calculate the $\chi^{2}$. Both profiles and obtained values of
$\chi^{2}$ are shown in Fig.~\ref{asymmetry}. The highest $\chi^{2}$
values were obtained for the P~Cygni profiles in ObsID~11044.
\ion{Mg}{xii} in ObsID~8525 and \ion{Ne}{x} in ObsID~3815 would be
also described as asymmetric, but for the latter this might also be
caused by blending with Fe lines. Setting a threshold of
$\chi^{2}=1.5$, all lines in ObsID~9847, \ion{Si}{xiv} and \ion{Ne}{x}
in ObsID~8525, \ion{Mg}{xii} in ObsID~3814 and \ion{Si}{xiv} in
ObsID~3815 would be considered symmetric. For \ion{Ne}{x} in
ObsIDs~9847 and 8525, the data have, however, high uncertainties, and
are therefore not very sensitive to detecting asymmetry.

\subsection{Properties of the X-ray emitting gas: density diagnostics
  of the \ion{Mg}{xi} He-like triplet}\label{sec:mgtriplet}

Due to the P Cygni profiles, the location of the absorber in
ObsID~11044 is difficult to constrain. This observation, however,
contains a strong He-like Mg triplet, i.e., the forbidden,
intercombination, and recombination lines f, i, and r, between 9.1 and
9.4\,\AA\ which is suitable for plasma diagnostics
(Fig.~\ref{11044-spectrum}). The existence of these lines allows us to
determine the physical condition of the location of the emitter,
although the absorber location is indeterminable. With the triplet
lines all being seen in emission, their intensity ratios
$G(T_\mathrm{e})=(\mathrm{i}+\mathrm{f})/\mathrm{r}$ and
$R(n_\mathrm{e})=\mathrm{f}/\mathrm{i}$ can be used to directly
estimate electron temperature and density of the emitting plasma
\citep{gabriel69a,porquet00a}. Note that in ObsID~3814,
$\phi_\mathrm{orb}\sim0.0$, which we studied in paper~I, only \ion{Mg}{xi}\,i
and f are seen in emission, while \ion{Mg}{xi}\,r is in absorption,
allowing only for the $R$-ratio to be analyzed.

In ObsID~11044, $G=1.6\pm0.5 \ll 4$ indicates a hybrid plasma where
next to photoionization, collisional ionization plays an important
role as well \citep{porquet00a}. The $G$ value corresponds to an
electron temperature of roughly (3--$5)\times10^{6}$\,K, depending on
the ratio of H-like to He-like ions. Since $G$ is a rather steep
function of temperature in that range, the uncertainty of $G$ has only
very little effect on the temperature. Note that the P~Cygni profile
of the resonance line also adds an uncertainty to its intensity, and
therefore, $G$.

The $R$-ratio of $0.6\pm0.5$ corresponds to an electron density of
$\sim$$4\times10^{13}\,\mathrm{cm}^{-3}$. The presence of the strong
UV field produced by HDE\,226868 can, however, deplete the metastable
upper level of the forbidden line by photoexcitation, lowering the
measured $R$-ratio \citep{mewe78a,kahn01a}. Thus, the derived density
constitutes an upper limit.

Similar results can be obtained by directly fitting the He-like
triplet using results from collisional plasma calculations by
D.~Huenemoerder (priv.\ comm., see
\url{http://space.mit.edu/cxc/analysis/hemodifier/index.html}), which
however, does not take the effects of UV radiaton into account either.
This fit gives $T_\mathrm{e}\sim3\times10^{6}$\,K and
$n_\mathrm{e}\sim6\times10^{13}\,\mathrm{cm}^{-3}$. 

In ObsID~3814 we could resolve two sets of \ion{Mg}{xi}~i and~f lines,
one unshifted and the other one redshifted (paper~I). Within their
uncertainties, the plasma seen in ObsID~11044 shows the same
$R$-ratio, i.e., the same electron density, as the redshifted Mg
triplet in ObsID~3814.

We note that the density determined from the emission lines is a
factor of $\sim$1000 higher than the density of the region where the
absorption is thought to occur. This could mean that the emission
component could originate in emitting high density regions, such as
clouds, that are mixed into the absorber. Alternatively, the high
density could point at an origin not in the wind but in the denser
regions of the accretion flow within the Bondi radius.

\section{Summary} \label{sec:summary}

Stellar winds of massive stars are highly inhomogenous phenomena: they
show temperatures spanning several orders of magnitude, turbulent
velocities, and highly variable densities. Gratings of the \textit{Chandra}
observatory provide us with the first real opportunity to investigate
the narrow spectral lines of the stellar winds in great detail and so
allow us to probe this variable environment. Cyg\,X-1 as a nearby, bright,
and persistent source is an ideal target for such a study. In this
work, we presented the most extensive analysis of the \textit{Chandra}-HETGS
hard state spectra of Cyg\,X-1 so far. We investigated five spectra that
cover the most important segments of orbital phase: stellar wind in
its focused stream at $\phi_\mathrm{orb}\sim0.0$, the opposite site of the
focused wind at $\phi_\mathrm{orb}\sim0.5$, as well as at parts perpendicular to
the binary axis at $\phi_\mathrm{orb}\sim0.2$ and $\sim$0.75. This is the second
paper in the series; paper~I presented analysis of one of the
observations (ObsID~3814, $\phi_\mathrm{orb}\sim0.95$), focusing only on spectra
extracted at the constant flux level (dips exluded). Here, we analyzed
the non-dip spectra of the remaining observations, and discussed their
variation over the orbital phase.

The most important results of our analysis are:
\begin{enumerate}
\item The intensity of dipping in light curves is phase dependent.
  Strong dips are observed along the binary axis at $\phi_\mathrm{orb}\sim0.0$
  while the light curve outside the focused stellar wind at
  $\phi_\mathrm{orb}\sim0.5$ is free of absorption dips in these
    observations.

\item Non-dip spectra reveal absorption lines of H- and He-like ions
  of S, Si, Al, Mg, Na, and Ne as well as L-shell Fe transitions
  corresponding to highly ionzed material. Spectra measured close to
  $\phi_\mathrm{orb}\sim0.0$ also show lines of lower ionized stages of
  \ion{Si}{xii}--\ion{Si}{ix}. The spectrum observed outside of the
  wind at $\phi_\mathrm{orb}\sim0.5$ shows clear P~Cygni profiles.

\item The neutral Ne K absorption edge was prominent in all
  investigated spectra, such that we were able to measure the neutral
  Ne column density independently from total column. The values vary
  moderately within
  $N_\mathrm{Ne}=(0.5$--$1.5)\times10^{18}\,\mathrm{cm}^{-2}$,
  attaining lowest values outside of the wind at
  $\phi_\mathrm{orb}\sim0.5$, and highest values at
  $\phi_\mathrm{orb}\sim0.2$. The Fe absorption edge was only resolved
  in the spectrum at $\phi_\mathrm{orb}\sim0.5$, giving
  $N_\mathrm{Fe}=0.15\pm0.02\times10^{18}\,\mathrm{cm}^{-2}$. Deriving
  $N_\mathrm{H}$ values from continuum fitting to the gratings data
  suffers from a large systematic error and yields values that are
  strongly biased towards lower columns.

\item Column densities obtained from the analysis of the ionized
  absorption lines are phase dependent, and therefore intrinsic to the
  source. They are highest around $\phi_\mathrm{orb}\sim0.0$, i.e., in the
  focused wind stream, and imply that the observed medium is almost
  fully ionized. The observed elements appear to be overabundant with
  respect to solar abundances.

\item Doppler shifts, $\pm500\,\mathrm{km}\,\mathrm{s}^{-1}$, vary
  strongly with orbital phase. Both, the column of the absorber and
  the Doppler shift are consistent with an absorber location that is
  less than $0.5d$ away from the black hole, where $d$ is the distance
  between the donor star and the black hole. The observations suggest
  that we observe a complex ionization structure.

\item The spectrum at $\phi_\mathrm{orb}\sim0.5$ shows the Mg He-triplet
  suitable for plasma diagnostics, suggesting that we also see a
  plasma with an electron temperature of
  $(3$--$5)\times10^{6}\,\mathrm{K}$ which is much denser than the
  locations in the wind responsible for absorption, e.g., clumps in
  the stellar wind.
\end{enumerate}
In the next paper of this series we will be studying the properties of
the gas responsible for the absorption dips in the light curves.

\begin{acknowledgements}
  The research leading to these results was funded by the European
  Community's Seventh Framework Programme (FP7/2007-2013) under grant
  agreement number ITN 215212 ``Black Hole Universe'' and by the
  Bundesministerium f\"ur Wirtschaft und Technologie under grant
  numbers \mbox{DLR 50\,OR\,0701} and \mbox{50\,OR\,1113}. This work
  was partially completed by Lawrence Livermore National Laboratory
  (LLNL) under Contract DE-AC52-07NA27344, and is supported by NASA
  grants to LLNL. Support for this work was also provided by NASA
  through the Smithsonian Astrophysical Observatory (SAO) contract
  SV3-73016 to MIT for Support of the Chandra X-Ray Center (CXC) and
  Science Instruments; CXC is operated by SAO for and on behalf of
  NASA under contract NAS8-03060. We acknowledge the support by the
  DFG Cluster of Excellence ``Origin and Structure of the Universe''
  and are grateful for the support by MCB through the Computational
  Center for Particles and Astrophysics (C2PAP). We are thankful to
  John E. Davis for the \texttt{SLXfig} package that was used to
  create the figures throughout this paper, to David Huenemoerder for
  the \textsl{AGLC} routine handling the grating lightcurves and
  routines related to plasma diagnostics, and to Thomas Dauser for
  parallel computing routines. This research has made use of the MAXI
  data provided by RIKEN, JAXA, and the MAXI team.
\end{acknowledgements}

\Online

\appendix
\section{Line Series Parameters \& Spectra}\label{sec:pars_and_spectra}

\begin{landscape}
\begin{table}
\caption{H-like, He-like and Fe absorption lines at
  $\phi_\mathrm{orb}\sim0.75$, $\phi_\mathrm{orb}\sim0.05$, and
  $\sim0.2$ (ObsID~3815, 8525, and 9847)}\label{abslines} 
\centering
\begin{tabular}{lccccccccc} 
\hline 
\hline
 & \multicolumn{3}{c}{ObsID 3815} & \multicolumn{3}{c}{ObsID 8525} & \multicolumn{3}{c}{ObsID 9847} \\
 & \multicolumn{3}{c}{$\phi_\mathrm{orb}\sim0.75$} & \multicolumn{3}{c}{$\phi_\mathrm{orb}\sim0.05$} & \multicolumn{3}{c}{$\phi_\mathrm{orb}\sim0.2$} \\
\hline
 & Column & Velocity & Thermal  & Column & Velocity & Thermal  & Column & Velocity & Thermal\\
 & density & shift & broadening & density & shift & broadening & density & shift & broadening\\
 & $N_{i}$ & $\Delta v$ & $\xi$ & $N_{i}$ & $\Delta v$ & $\xi$ & $N_{i}$ & $\Delta v$ & $\xi$\\
 & [$10^{16}$\,$\mathrm{cm}^{-2}$] & [$\mathrm{km}\,\mathrm{s}^{-1}$] & [$\mathrm{km}\,\mathrm{s}^{-1}$]  & [$10^{16}$\,$\mathrm{cm}^{-2}$] & [$\mathrm{km}\,\mathrm{s}^{-1}$] & [$\mathrm{km}\,\mathrm{s}^{-1}$] & [$10^{16}$\,$\mathrm{cm}^{-2}$] & [$\mathrm{km}\,\mathrm{s}^{-1}$] & [$\mathrm{km}\,\mathrm{s}^{-1}$]\\
\hline
\multicolumn{10}{l}{H-like lines}\\
\hline
\ion{Ne}{x} & $1.76^{+0.16}_{-0.13}$ & $190^{+27}_{-22}$ & $240\pm50$ & $160^{+130}_{-70}$& $-130^{+70}_{-60}$& $73^{+20}_{-18}$& $160^{+80}_{-60}$ & $-360\pm40$ & $69^{+16}_{-12}$\\
\ion{Na}{xi} & $1.94^{+0.18}_{-0.17}$ & $350^{+100}_{-90}$ & $1120^{+140}_{-120}$ & $7.3^{+1.7}_{-1.6}$& $0.0^{+320}_{-260}$& $1400^{+600}_{-400}$& $6.9^{+1.3}_{-0.9}$& $240^{+360}_{-220}$ & $1910^{+500}_{-270}$ \\
\ion{Mg}{xii} & $3.24\pm0.15$ & $311^{+20}_{-18}$ & $328^{+30}_{-29}$ & $8.1^{+1.9}_{-0.7}$& $-80\pm70$& $370^{+130}_{-100}$& $6.1^{+1.0}_{-1.3}$ & $-200\pm90$ & $420^{+90}_{-250}$\\
\ion{Al}{xiii} &  $0.61^{+0.14}_{-0.13}$ & $130\pm100$ & $360^{+240}_{-180}$ & $\le7.9$ & $0.0\pm10^{4}$ & $\le3\times10^{4}$& $\le8.2$ & $\left(-0.06^{+1.05}_{-0.20}\right)\times10^{4}$ & $\le8540$\\
\ion{Si}{xiv} &  $4.44^{+0.20}_{-0.19}$ & $259^{+27}_{-22}$ & $420\pm50$ & $300^{+100}_{-80}$& $-90\pm50$& $\le517$& $6.7^{+271}_{-1.4}$ & $-180^{+100}_{-80}$ & $200^{+200}_{-140}$\\
\ion{S}{xvi} & $3.4\pm0.5$ & $230^{+80}_{-70}$ & $220\pm170$ & $6.1^{+173.2}_{-2.2}$& $-100\pm400$& $320^{+490}_{-280}$& $60^{+40}_{-50}$ & $-323.64^{+0.24}_{-120.51}$ & $\le2.8$\\
\ion{Ar}{xviii} & \multicolumn{3}{c}{--} &  $5\pm4$& $\left(-1.00^{+0.08}_{-0.00}\right)\times10^{4}$ & $1400^{+1900}_{-1200}$ & $\le26$ & $\left(0.1\pm1.0\right)\times10^{4}$ & $\le1875$\\
\ion{Ca}{xx} & \multicolumn{3}{c}{--} & $14^{+76}_{-13}$ & $-400^{+600}_{-1000}$& $\le2586$ & $\le60$ & $\le10^{4}$ & $\le6800$ \\
\hline
\multicolumn{10}{l}{He-like lines}\\
\hline
\ion{Ne}{ix} & $0.89\pm0.13$ & $390^{+100}_{-90}$ & $650^{+140}_{-100}$ & $19^{+12}_{-9}$ & $150^{+90}_{-120}$& $120^{+140}_{-50}$& $10.5^{+2.8}_{-2.3}$ & $-250^{+130}_{-150}$ & $410^{+120}_{-100}$\\ 
\ion{Na}{x} & $0.99\pm0.09$ & $870^{+90}_{-70}$  & $830^{+100}_{-80}$ & $6.0^{+138.5}_{-2.8}$& $-300\pm100$& $\le130$& $50^{+35}_{-25}$ & $220\pm60$ & $8.9^{+6.7}_{-1.8}$\\
\ion{Mg}{xi} &  $0.68\pm0.08$ & $370^{+60}_{-50}$ & $350^{+80}_{-90}$  & $11^{+6}_{-4}$& $-160^{+80}_{-70}$& $84^{+37}_{-25}$& $10^{+8}_{-4}$ & $-200^{+80}_{-70}$ & $76^{+29}_{-24}$\\
\ion{Al}{xii} & $0.22^{+0.06}_{-0.07}$ & $240^{+120}_{-100}$ & $270^{+190}_{-260}$ & $7.2^{+2.2}_{-2.3}$& $\left(1.00^{+0.00}_{-0.26}\right)\times10^{4}$\,\tablefootmark{\dag} & $\left(2.3^{+0.7}_{-0.6}\right)\times10^{4}$\,\tablefootmark{\dag} & $7^{+8}_{-6}$ & $-150^{+80}_{-160}$ & $11^{+17}_{-10}$\\
\ion{Si}{xiii} &  $4.4^{+1.0}_{-1.9}$ & $317^{+1}_{-40}$ & $16^{+7}_{-5}$ & $14^{+14}_{-7}$& $-210^{+90}_{-50}$& $93^{+33}_{-25}$& $\le90$ & $-310^{+150}_{-60}$ & $220^{+95}_{-23}$\\
\ion{S}{xv} &  $1.37^{+0.22}_{-0.20}$ & $230^{+80}_{-70}$ & $310\pm150$  & $4.2\pm1.5$& $300\pm400$& $900\pm600$& $\le37$ & $-30^{+80}_{-160}$ & $16^{+63}_{-6}$\\
\ion{Ar}{xvii} &  $\le0.5$ & $\left(0.04^{+0.96}_{-0.20}\right)\times10^{4}$\,\tablefootmark{\dag} & $\left(2.25^{+0.75}_{-0.04}\right)\times10^{4}$\,\tablefootmark{\dag} & \multicolumn{3}{c}{--} & $3.7297^{+0.0004}_{-2.4044}$ & $\left(0.766^{+0.001}_{-0.008}\right)\times10^{4}$ & $36.968^{+0.004}_{-27.427}$\\
\ion{Ca}{xix} & \multicolumn{3}{c}{--} & \multicolumn{3}{c}{--} & $\le8$ & $\le10^{4}$\,\tablefootmark{\dag} & $4300^{+2700}_{-3500}$\\
\hline
\multicolumn{10}{l}{Fe lines}\\
\hline
\ion{Fe}{xvii} & \multicolumn{3}{c}{--} & $3^{+73}_{-1}$ & $-260^{+140}_{-130}$ & $130^{+120}_{-100}$ & $3^{+23}_{-1}$& $-230^{+90}_{-80}$& $80^{+90}_{-70}$\\
\ion{Fe}{xviii} & $1.47\pm0.16$ & $144^{+28}_{-18}$ & $34^{+7}_{-3}$ & $4.2^{+1.5}_{-1.3}$ & $-210^{+100}_{-120}$ & $100^{+130}_{-50}$ & $3.3^{+1.7}_{-1.0}$& $-410^{+130}_{-140}$& $100^{+100}_{-60}$\\
\ion{Fe}{xix} & $2.96^{+0.29}_{-0.23}$ & $225.2\pm0.2$ & $19.06^{+0.89}_{-0.07}$  & $6.7^{+1.5}_{-1.3}$ & $-390^{+100}_{-90}$ & $150^{+130}_{-60}$ & $2.3^{+0.9}_{-0.7}$& $-220^{+160}_{-200}$& $130^{+310}_{-80}$\\
\ion{Fe}{xx} & $1.64^{+0.10}_{-0.12}$ & $60\pm40$ & $400^{+40}_{-70}$ & $4.6^{+1.8}_{-1.0}$ & $-130^{+190}_{-200}$ & $350^{+370}_{-220}$ & $3.8^{+1.7}_{-1.0}$& $-330^{+150}_{-140}$& $200^{+260}_{-100}$\\
\ion{Fe}{xxi} & $1.16^{+0.09}_{-0.08}$ & $240\pm50$ & $740^{+80}_{-70}$ & $2.7^{+0.9}_{-0.8}$ & $-310^{+250}_{-200}$ & $420^{+280}_{-200}$ & $1.6\pm0.4$& $-440^{+130}_{-220}$& $220^{+350}_{-50}$\\
\ion{Fe}{xxii} & $2.39^{+0.36}_{-0.28}$ & $182\pm10$ & $21\pm3$ & $2.6^{+1.0}_{-0.8}$ & $100^{+190}_{-250}$ & $230^{+350}_{-100}$ & $7.3^{+3.0}_{-4.9}$& $-158^{+1}_{-3}$& $12^{+32}_{-6}$\\
\ion{Fe}{xxiii} & $6.2^{+1.0}_{-2.6}$ & $231^{+10}_{-9}$ & $8.4^{+3.4}_{-1.6}$ & $\le34$ & $200^{+500}_{-1000}$ & $\le864$ & $\le1.6\times10^{-3}$& $\left(0.0\pm1.0\right)\times10^{4}$\,\tablefootmark{\dag} & $\le1.27\times10^{4}$\\
\ion{Fe}{xxiv} & $2.21^{+0.18}_{-0.13}$ & $261^{+25}_{-15}$ & $250\pm50$ & $70^{+100}_{-70}$ & $-114^{+26}_{-64}$ & $10^{+10}_{-6}$ & $4.5\pm1.0$& $10\pm60$& $440^{+230}_{-190}$\\
\hline
\hline 
\end{tabular}
\tablefoot{\tablefoottext{\dag}{Unconstrained parameter. The calculation has reached one or both limits of the range set for this parameter. Note that primarily affected line series are those consisting of weak or possibly blended lines of Al, Ar, Ca or Fe.}
}
\end{table}
\end{landscape}

\begin{landscape}
\begin{table}
\caption{Parameters for spectral line features for ObsID~11044.
All lines, including the emission and absorption line portion of the
P~Cygni line profiles, are modeled with Voigt
profiles.}\label{11044_p-cygni_my} 
\centering
\begin{tabular}{lc cccccc ccccc}
\multicolumn{12}{l}{\large{P~Cygni profiles}} \\
\hline
\hline
  & & \multicolumn{6}{l}{Absorption} & \multicolumn{4}{l}{Emission} \\
  \hline

  & Reference  & Obs.   &      & Thermal
  &      & Velocity &
  Column  & Obs.   & Norm &
  Thermal    &      & Velocity\\
  & wavelength & wavel. & Norm &
  broadening & FWHM & shift    &
  density &  wavel. &      & broadening & FWHM & shift\\
  Line & $\lambda_\mathrm{theory}$ & $\lambda_\mathrm{obs}$ & $A$ & $v_\mathrm{therm}$ & $\Gamma$ & $\Delta v$ & $N_i$  
       & $\lambda_\mathrm{obs}$ & $A$ & $v_\mathrm{therm}$ & $\Gamma$
       & $\Delta v$ \\ 
       & (\AA) & (\AA) & ($10^{-3}\,\mathrm{ph}\,\mathrm{s}^{-1}\,\mathrm{cm}^{2}$) & ($\mathrm{km}\,\mathrm{s}^{-1}$) & (keV) & ($\mathrm{km}\,\mathrm{s}^{-1}$) & ($10^{16}\,\mathrm{cm}^{-2}$) 
       & (\AA) & ($10^{-3}\,\mathrm{ph}\,\mathrm{s}^{-1}\,\mathrm{cm}^{2}$) & ($\mathrm{km}\,\mathrm{s}^{-1}$) & (keV) & ($\mathrm{km}\,\mathrm{s}^{-1}$) \\
  \hline
\ion{Si}{xiv} Ly$\alpha$  & 6.1805 & $6.166(2)$ & $-0.12\pm0.05$ & $90^{+200}_{-80}$ & -- & $-700\pm100$ & $0.9(3)$ & 6.181(3) & $0.21\pm0.06$ & $310^{+200}_{-230}$ & -- & $20\pm150$\\
\ion{Mg}{xii} Ly$\alpha$ & 8.4192 & $8.402(4)$ & $-0.13\pm0.07$  & $\le200$ & $\le0.010$ & $-610\pm140$ & $0.7(2)$ & 8.419(3) & $0.17^{+0.07}_{-0.05}$  & $270^{+220}_{-190}$ & -- & $-10\pm110$  \\
\ion{Ne}{x} Ly$\alpha$ & 12.1317 & $12.115(3)$ & $-0.7^{+0.2}_{-0.1}$  &$500^{+140}_{-340}$ & $\le0.012$ & $-410\pm70$ & $3.0(4)$ & 12.128(2) & $0.97^{+0.35}_{-0.28}$  & $\le269$ & $0.009^{+0.005}_{-0.008}$ & $-90\pm50$ \\
\ion{Ne}{x} Ly$\beta$ & 10.2327 & $10.231(2)$ & $-4.2^{+2.6}_{-2.3}$&$700^{+800}_{-500}$ & $0.04^{+0.01}_{-0.03}$ & $-50\pm60$ & $74(2)$ & 10.237(2) & $4.8^{+2.4}_{-2.7}$ & $\le1082$ & $0.056^{+0.014}_{-0.037}$ & $130\pm60$ \\
\ion{Ne}{x} Ly$\delta$ & 9.4794 & $9.464(1)$ & $-0.56^{+0.22}_{-0.38}$ & $2000^{+1900}_{-1600}$& $\le0.05$ & $-490\pm30$ & $83(19)$ & 9.500(6) & $0.27^{+0.36}_{-0.16}$ & $\le821$ & $\le0.05$ & $650\pm190$\\
\\
\multicolumn{12}{l}{\large{\ion{Mg}{xi} triplet}} \\
\hline
\hline
\ion{Mg}{xi} r & 9.1687 & $9.14(1)$ & $-0.3\pm0.1$ & $\le200$ & $0.05^{+0.00}_{-0.02}$ & $-940\pm330$ & $0.9(4)$ & 9.168(2) & $0.26^{+0.06}_{-0.07}$ &  $280\pm100$ &  $\le0.008$ & $-20\pm70$ \\
\ion{Mg}{xi} i & 9.2312 &    &                   &          &                      & & & 9.228(5) & $0.22^{+0.16}_{-0.06}$ & $\le700$ & $\le0.026$ & $-100\pm160$ \\
\ion{Mg}{xi} f & 9.3143 &    &                   &          &                     & & & 9.308(7) & $0.16^{+0.07}_{-0.09}$ & $\le810$ & $\le0.025$ & $-200\pm230$ \\
\\
\multicolumn{12}{l}{\large{Emission lines}} \\
\hline
\hline
Fe\,K$\alpha$ & 1.9373 & & & & & & & 1.937(3) & $1.0\pm0.4$& $\le950$& $\le0.5$ & $-50\pm450$\\
\ion{Fe}{xxv} & 1.8504 & & & & & & & 1.862(6) & $0.22\pm0.2$& $\le970$& $\le0.5$ & $1880\pm970$\\
\ion{Si}{xiii} He\,f & 6.7403 & & & & & & & 6.741(1) & $0.28\pm0.07$& $\le240$& $\le0.017$ & $30\pm40$\\
\ion{Ne}{ix} He\,i & 13.55 & & & & & & & 13.55(2) & $0.28\pm0.12$& $600\pm400$& -- & $0\pm440$ \\
\\
\multicolumn{12}{l}{\large{Absorption lines}} \\
\hline
\hline
\ion{Si}{xii} (Li) & 6.7176 & $6.705(1)$ & $-0.13\pm0.04$ & $26^{+27}_{-8}$ & -- & $-670\pm50$ & $0.6(2)$ & & & & \\
\ion{Fe}{xxi} & 12.2840 & $12.265(7)$ & $-0.15\pm0.06$ & $170^{+180}_{-160}$ & -- & $-460\pm170$ & -- & & & & \\
\hline  
\hline
 \end{tabular}
\tablefoot{Number in brackets corresponds to the uncertainty in units
  of the last digit. Equivalent width and column density are not
  calculated for emission lines.} 
\end{table}
\end{landscape}

\begin{figure*}\centering
\centering\includegraphics[width=17cm]{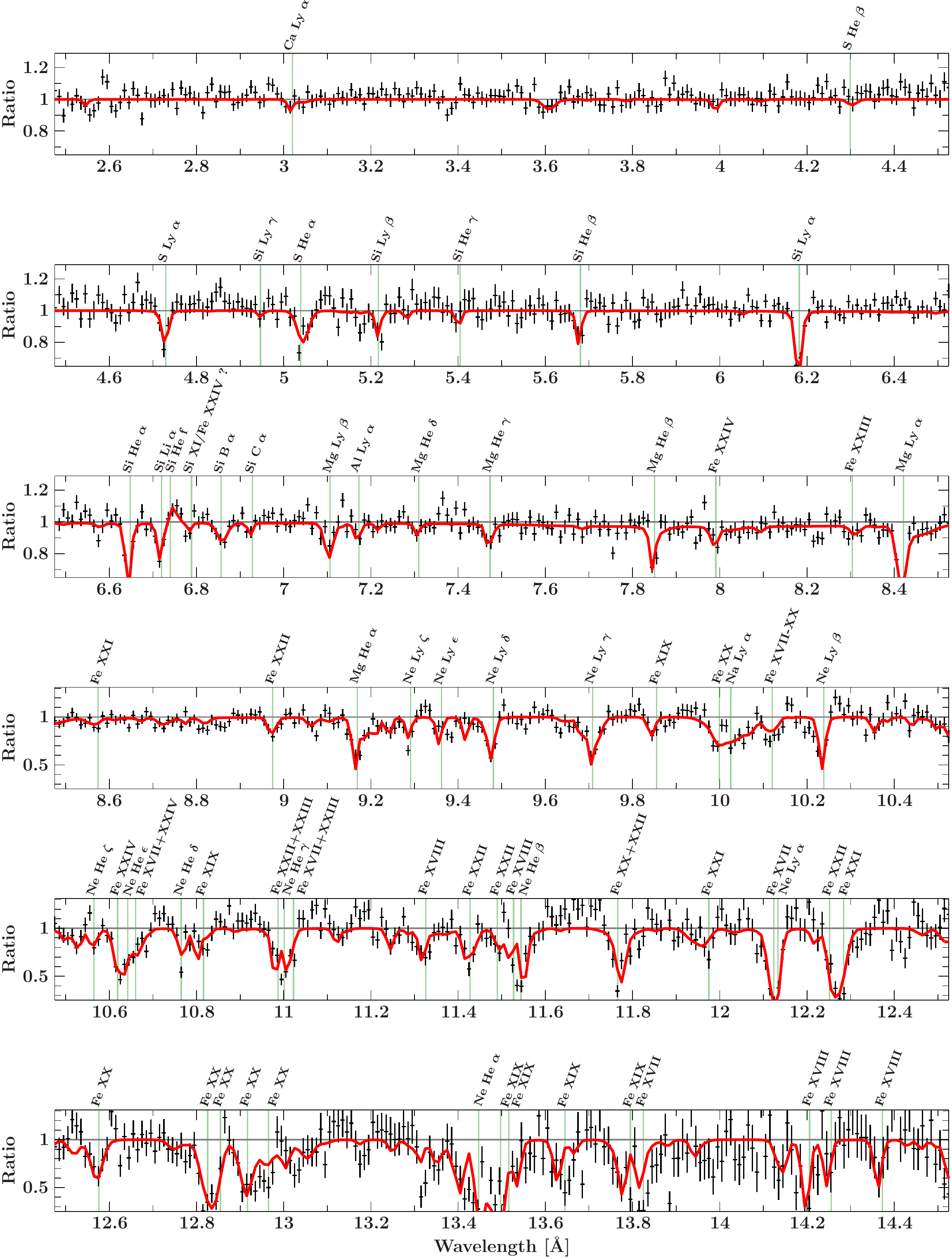}
\caption{Non-dip spectrum of ObsID~8525 -- displayed as the ratio
  between data and absorbed power-law continuum model -- showing the
  absorption lines observed in the spectrum. Due to a relatively short
  exposure time of only 4.4\,ks, the HEG spectra are worse than MEG at
  longer wavelengths. Data from HEG were therefore scaled to MEG data
  for visual reasons. The red line shows lines modeled with line
  series. Each of them is labeled at its rest wavelength. There
  are no significant emission lines in the spectrum except the
  \ion{Si}{xiii}\,forbidden line. Instead, absorption lines from
  \ion{Si}{xii} (Li-like), possibly \ion{Si}{xi} (Be-like),
  \ion{Si}{x} (B-like), and \ion{Si}{ix} (C-like), although
  quite weak, can be clearly identified in the ``Si region'' between
  6.6\,\AA\ and 7\,\AA.
  \label{8525_spectrum}}
\end{figure*}
 
\begin{figure*}\centering
\centering\includegraphics[width=17cm]{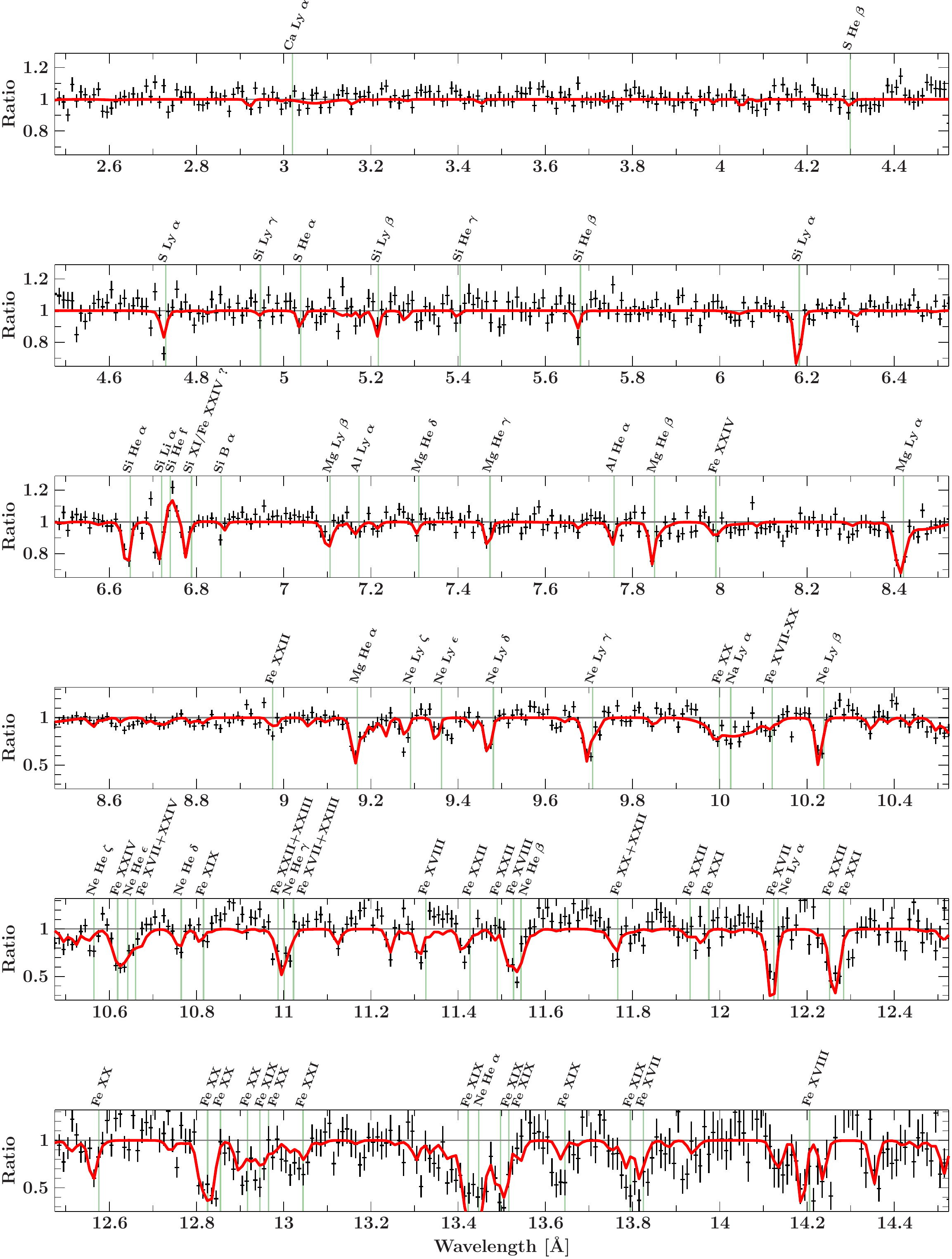}
\caption{Line identifications for ObsID~9847. Non-dip data displayed as ratio between data and
  absorbed power-law continuum model.\label{9847-spectrum}}
\end{figure*}

\begin{figure*}\centering
\includegraphics[width=17cm]{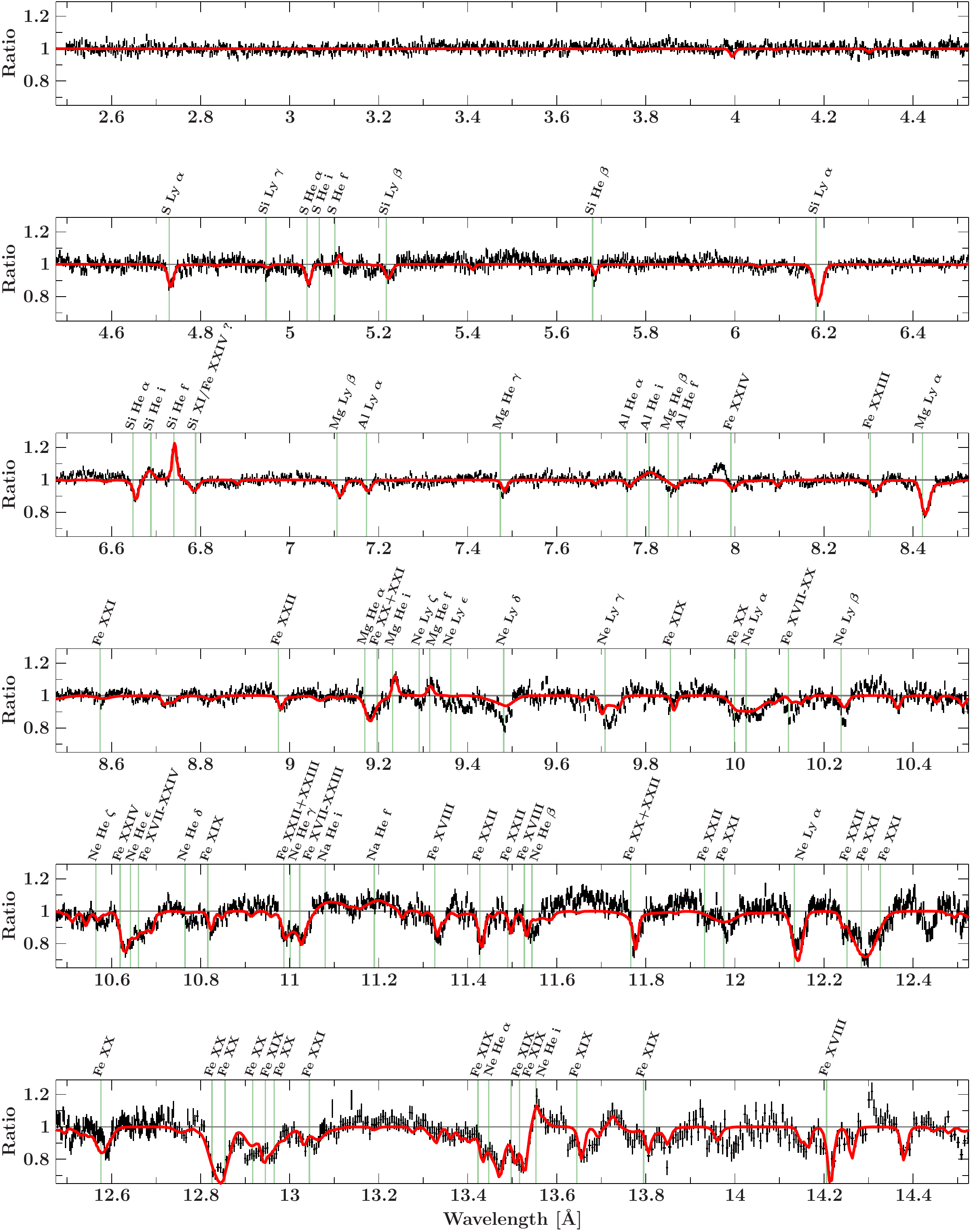}
\caption{Non-dip spectrum of ObsID~3815 -- displayed as the 
  ratio between data and absorbed power-law continuum model -- showing
  the H- and He-like absorption lines, as well as the i and f emission
  lines of the hot gas of the stellar wind. The spectra of HEG$\pm$1
  and MEG$\pm$1 were combined and binned to the same signal-to-noise
  ratio ($S/N=10$) as for fitting.  Especially notable are lines of
  \ion{Ne}{x}$\beta$ at 10.24\,\AA, \ion{Ne}{x}$\gamma$ at 9.71\,\AA\ and
  \ion{Ne}{x}$\delta$ at 9.48\,\AA, because they are much stronger
  in the data than the profiles that are predicted by the model. We
  did not manage to identify the emission feature at $\sim$7.98\AA. It
  is visible also in ObsID~1044 (Fig.~\ref{11044-spectrum}), but not
  in ObsIDs~8525 and 9847 (Figs.~\ref{8525_spectrum} and~\ref{9847-spectrum}).
  \label{abs-lines}}
\end{figure*}
 
\begin{figure*}\centering
\includegraphics[width=17cm]{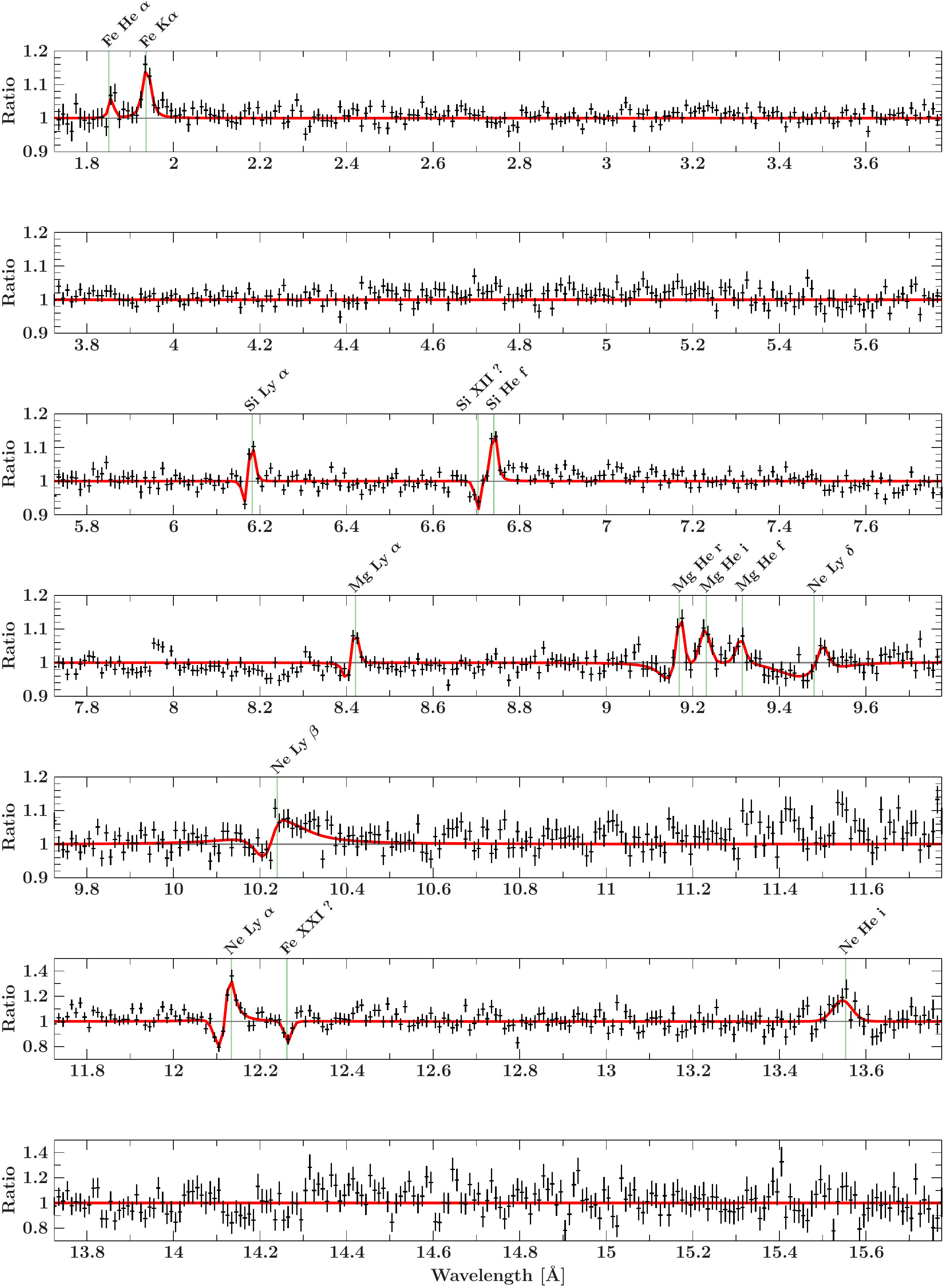}
\caption{ Spectrum of ObsID~11044 displayed as the ratio between data and
  absorbed power-law continuum model. The most prominent absorption
  and emission lines have been fitted with Voigt profiles. The Ly$\alpha$
  lines of H-like Si, Mg, and Ne show clear P~Cygni profiles. The
  absorption line at 6.70\,\AA, however, cannot be related to the
  emission line at 6.74\,\AA, if the latter is due to the (dipole-)
  forbidden transition of \ion{Si}{xiii}. 
  \label{11044-spectrum}}
\end{figure*} 

\clearpage

\section{Diffraction Order Determination in CIAO Data Reduction}\label{sec:order-sorting}

\begin{figure*}\centering
\includegraphics[width=17cm]{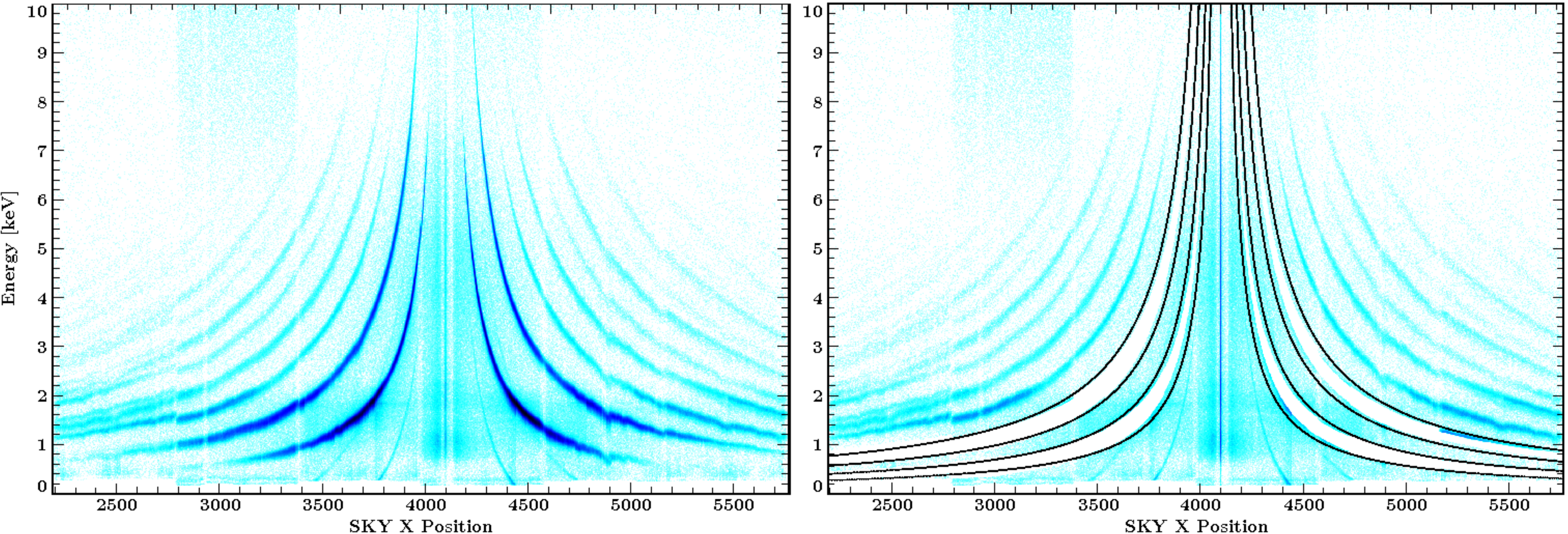}
\caption{ ``Banana'' plot -- order-sorting of incoming events.
  \textbf{Left:} Diffraction orders for the HEG and MEG instruments.
  The source itself is located at sky \textsl{x}-position
  \textsl{x}=4100. The higher density of events around this region is
  most likely associated with a scattering halo. The different orders
  correspond to hyperbolas, shaped by a higher density of events. The
  hyperbolas with the highest event density, close to the source,
  represent the $\pm1^\mathrm{st}$ order of HEG and MEG.
  \textbf{Right:} Same as left figure, but showing the events assigned
  to the HEG and MEG in standard processing (white regions). Not
  all of the events are included. Changing the OSIP parameters as
  described in the text allows to include all events belonging to each
  order in the analysis (black lines).
\label{banana}}
\end{figure*}

A data processing problem that would have strongly affected any
further analysis consists in the incorrect determination of the
diffraction order of photons in \textit{Chandra}'s standard data reduction
pipelines. Here, based on detector (ACIS) energy and position
(dispersion distance or wavelength) of X-ray photons, it is possible
to disentagle events into different spectral orders, as shown in
so-called ``banana'' plots (Fig.~\ref{banana}). Different event
hyperbolas correspond to individual orders (left panel), but
especially in observations of bright sources it can happen that not
all events are assigned to their associated order, since the charge
transfer efficiency in the detectors changes with source flux. This so
called order-sorting problem is most prominent in the first order
spectra (right panel). This problem can be solved by manually changing
the default values in the Order Sorting and Integrated Probability
(OSIP) file used by the \texttt{tg-resolve-events} routine of the
standard data processing and ignoring the default values obtained from
\textit{Chandra}'s calibration database. Such a correction results in a much
wider range of events entering the analysis (Fig.~\ref{banana}).

\section{Telemetry Saturation}\label{sec:telemetry}

\begin{figure}
\resizebox{\hsize}{!}{\includegraphics{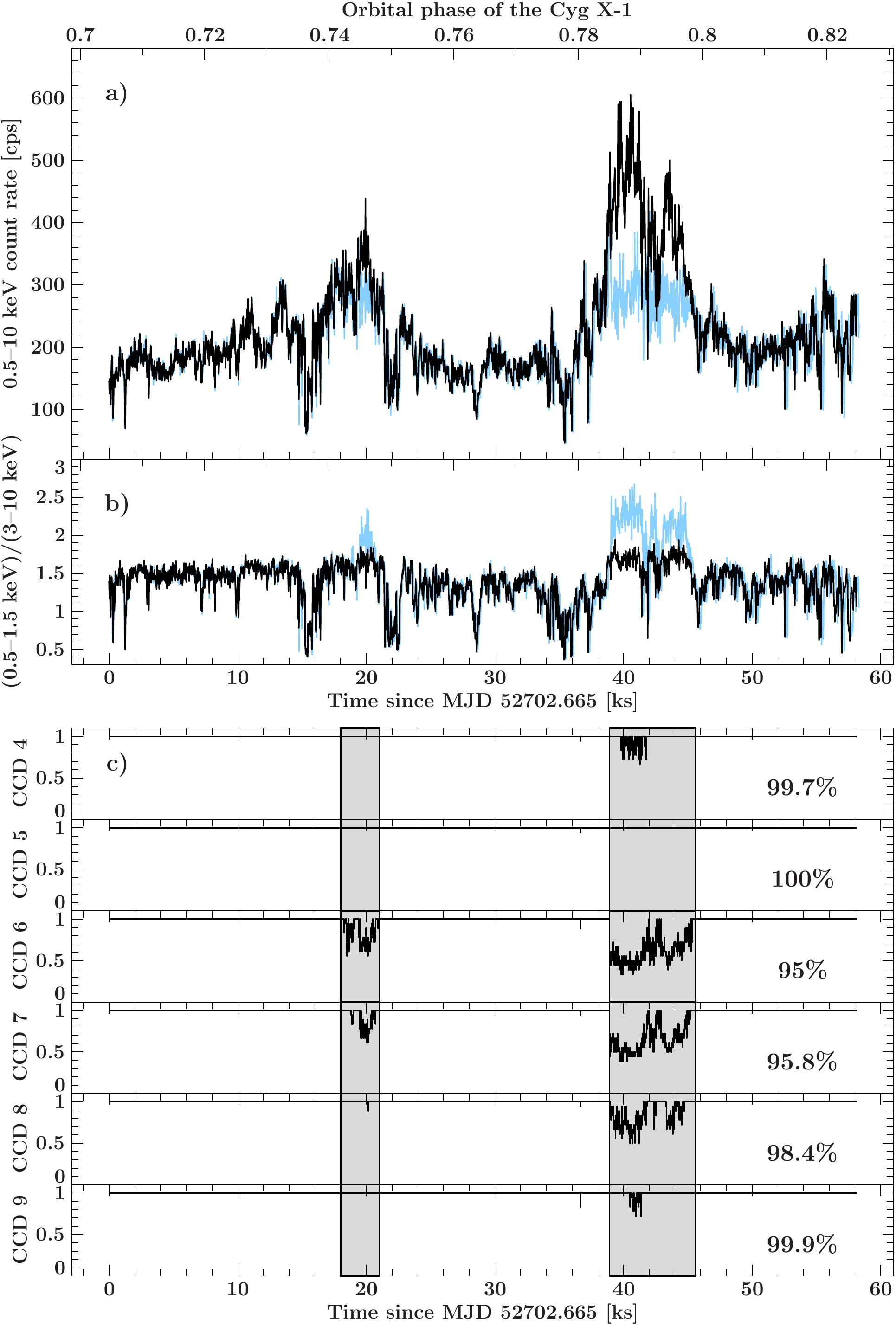}}
\caption{\textbf{a} 0.5--10\,keV \textit{Chandra} light curve of Cyg\,X-1 (time
  resolution of 25.5\,s), obtained in 2003 March (ObsID~3815). 
  Light blue: light curve without taking telemetry drop-outs into
  account. Black: light curve after correcting for telemetry
  drop-outs. \textbf{b} Ratio between the 0.5-1.5\,keV and 3-10\,keV
  energy bands, again with and without correction for the telemetry
  saturation. \textbf{c} Fractional exposure during the observation
  for each CCD. Telemetry drop-outs appeared at $\sim$20\,ks and after
  $\sim$40\,ks (gray regions). In the latter case, 5 of 6 CCDs were
  affected. Since whole data frames were discarded, lower countrates
  were observed in the light curve in the affected time intervals
  before the correction was applied (panel~a). The spectrum extracted
  for  those GTIs might be influenced as well, especially in the
  wavelength range of affected CCDs. These GTIs were not used in the
  analysis, even after corretion of the count rates.}\label{lc}
\end{figure}

A second problem occurring in sources with high flux is that because
of the large count rates data are generated so fast that the telemetry
stream may become saturated, causing an exposure gap in the
corresponding time bin. The remaining data frame from the affected CCD
chip is discarded (Fig.~\ref{lc}c). Count rate light curves from
observations affected by this problem -- in our ase ObsID~3815 -- will
erroneously show lower fluxes because they are commonly based on the
average CCD exposure instead of treating countrates for each CCD
separately. This problem can be dealt with by including exposure
statistics from the secondary Chandra data in the Chandra standard
processing using the tool \texttt{aglc} (D. Huenemoerder, priv.\ comm.;
see Fig.~\ref{lc}).

\end{document}